\documentclass[%
 reprint,
superscriptaddress,
 amsmath,amssymb,
 aps,
]{revtex4-2}

\usepackage{graphicx}
\usepackage{dcolumn}
\usepackage{bm}
\usepackage{bbold}
\DeclareMathOperator{\sgn}{sgn}

\usepackage{physics}
\usepackage{hyperref}
\hypersetup{colorlinks=true,linkcolor=blue,citecolor=blue,filecolor=blue,urlcolor=blue}

\begin{document}


\title{Exploring the Fidelity of Flux Qubit Measurement in Different Bases via the Quantum Flux Parametron}

\author{Yanjun Ji}%
 \email{y.ji@fz-juelich.de}
\affiliation{Institute for Quantum Computing Analytics (PGI-12), Forschungszentrum Jülich, 52425 Jülich, Germany}

\author{Susanna Kirchhoff}
\affiliation{Institute for Quantum Computing Analytics (PGI-12), Forschungszentrum Jülich, 52425 Jülich, Germany}
\affiliation{%
Theoretical Physics, Saarland University, 66123 Saarbrücken, Germany
}%

\author{Frank K. Wilhelm}
\email{f.wilhelm-mauch@fz-juelich.de}
\affiliation{Institute for Quantum Computing Analytics (PGI-12), Forschungszentrum Jülich, 52425 Jülich, Germany}
\affiliation{%
Theoretical Physics, Saarland University, 66123 Saarbrücken, Germany
}%

\date{\today}

\begin{abstract}

High-fidelity qubit readout is a fundamental requirement for practical quantum computing systems. In this work, we investigate methods to enhance the measurement fidelity of flux qubits via a quantum flux parametron-mediated readout scheme. Through theoretical modeling and numerical simulations, we analyze the impact of different measurement bases on fidelity in single-qubit and coupled two-qubit systems. For single-qubit systems, we show that energy bases consistently outperform flux bases in achieving higher fidelity. In coupled two-qubit systems, we explore two measurement models: sequential and simultaneous measurements, both aimed at reading out a single target qubit. Our results indicate that the highest fidelity can be achieved either by performing sequential measurement in a dressed basis over a longer duration or by conducting simultaneous measurement in a bare basis over a shorter duration. Importantly, the sequential measurement model consistently yields more robust and higher fidelity readouts compared to the simultaneous approach. These findings quantify achievable fidelities and provide valuable guidance for optimizing measurement protocols in emerging quantum computing architectures.

\end{abstract}

\maketitle


\section{Introduction}

Quantum computers have the potential to efficiently solve computational problems that are intractable for classical computers. Among the various physical implementations, superconducting qubits \cite{clarke2008superconducting, krantz2019quantum} stand out due to their scalability and compatibility with existing semiconductor fabrication techniques.
Josephson junctions, composed of two superconductors separated by a thin insulating barrier, are essential components in constructing diverse qubit types, including charge \cite{bouchiat1998quantum, nakamura1997spectroscopy}, phase \cite{martinis2009superconducting}, flux \cite{friedman2000quantum, van2000quantum}, and transmon qubits \cite{koch2007charge}. Flux qubits, in particular, show promise for quantum information processing due to their long coherence times \cite{stern2014flux}.
Furthermore, accurate and reliable qubit readout is crucial for practical quantum technologies. The properties and readout schemes of flux qubits have been extensively explored~\cite{schondorf2020flux, harris2010experimental, ferber2010efficient, fowler2007long, robertson2005superconducting, chiorescu2003coherent}. A widely used approach is dispersive measurement, which involves coupling superconducting qubits to microwave resonators \cite{blais2004cavity} and has been experimentally validated~\cite{walter2017rapid, heinsoo2018rapid, lin2014josephson}.
Additionally, the quantum flux parametron (QFP) \cite{hosoya1991quantum} has emerged as a notable candidate for qubit readout due to its high sensitivity and low power consumption \cite{yamanashi2017evaluation}. Its application as an isolator and amplifier within the readout circuit has been shown, enabling fast and high fidelity measurements~\cite{grover2020fast}. Despite these advancements, measurement fidelity remains influenced by various factors, with the choice of the appropriate measurement basis being a critical consideration \cite{pommerening2020measured, khezri2015qubit}.

The measurement postulate in its simplest form states that the measurement basis is determined by the detector \cite{braginsky1992quantum}. In practice, this is often realized only if this basis coincides with the energy eigenbasis of the measured system, leading to a quantum nondemolition (QND) measurement. However, if the measurement basis does not coincide with the energy eigenbasis and the measurement is system-dominated, spurious transitions occur, and the measurement loses contrast. Conversely, the regime where the detector dictates the measurement basis is referred to as a detector-dominated scenario. Specifically for mid-anneal measurements in adiabatic quantum computing, it would be highly desirable to measure flux qubits in a detector-dominated scenario \cite{albash2018adiabatic}.

In this work, we propose a QFP-mediated measurement scheme and identify the measurement basis yielding the closest projection. Through theoretical analysis and numerical simulations, we evaluate the performance of different measurement bases, including flux, energy, bare, and dressed bases, in both single and coupled qubit systems. Additionally, we compare sequential and simultaneous measurements in coupled systems to determine their effectiveness in reading out a single target qubit. Our comprehensive study provides deeper insight into measurement processes and highlights new scenarios that extend beyond the conventional system-dominated regime. This optimization advances progress toward achieving high-fidelity measurements of flux qubits, which is crucial for developing large-scale quantum computers. Moreover, our findings have the potential to enhance measurement fidelity in near-term quantum devices \cite{preskill2018quantum} and improve the resilience of quantum algorithms \cite{bharti2022noisy,ji2024resilience}, paving the way for more reliable quantum information processing.

This paper is organized as follows. Section~\ref{sec:prelim} provides background on QFPs, dispersive readout of qubits, and the measurement principle. Section~\ref{sec:analy} focuses on the measurement of a single flux qubit using a QFP-mediated readout scheme, comparing fidelity across different measurement bases. Section~\ref{sec:measu} extends this analysis to coupled flux qubit systems, examining both sequential and simultaneous measurement strategies. Section~\ref{sec:discu} discusses practical integration constraints, decoherence, and noise robustness. Finally, Sec.~\ref{sec:conclu} concludes the paper.

\section{Preliminaries \label{sec:prelim}}

\subsection{Quantum flux parametron \label{subsec:qfp}}

The QFP is a controllable rf-SQUID (radio-frequency superconducting quantum interference device) \cite{clarke2004squid} characterized by a small inductance, large capacitance, and a significantly higher critical current compared to a flux qubit \cite{hosoya1991quantum}. This configuration results in a larger tunnel barrier, suppressing tunnel coupling and enhancing state stability against high-frequency radiation during readout \cite{berkley2010scalable}.
Integrating a QFP between a flux qubit and its resonator offers two key advantages \cite{grover2020fast}: (1) amplification of the current signal, enabling more sensitive readout of qubit state; and (2) isolation of qubit from the resonator, effectively suppressing crosstalk that can degrade measurement accuracy.

The Hamiltonian describing the QFP and its interaction with a coupled flux qubit is given by \cite{harris2009compound, berkley2010scalable}
\begin{equation*}
\hat{H}_{\text{qfp}}=\frac{\hat{Q}^2}{2C}+\frac{\left(\hat{\Phi}-\Phi_{z}^{\text{qfp}}\right)^2}{2L}-E_J 
\cos\left(\frac{\pi\Phi_{x}^{\text{qfp}}}{\Phi_0}\right)\cos(\frac{2 \pi \hat{\Phi}}{\Phi_0}),
\end{equation*}
where $\hat{Q}$ is canonical conjugate to $\hat{\Phi}$, $C$ is the sum of two junction capacitances, and $L$ is the inductance of the large loop. The magnetic flux induced by the flux qubit, denoted as $\Phi_{z}^{\text{qfp}}$, is given by $\Phi_{z}^{\text{qfp}} = M_q I_p^{q} = M_q \abs{I_p^{q}} \hat{\sigma}_z$, where $M_q$ is the mutual inductance between the flux qubit and QFP, often referred to as coupling inductance, and $I_p^{q}$ is the persistent current of qubit, which takes values $\pm \abs{I_p^{q}}$ depending on the qubit state. $\hat{\sigma}_z$ is the Pauli Z operator, which represents a qubit state in the computational basis, with eigenvalues of $\pm 1$ corresponding to two persistent current states of the flux qubit. $\Phi_0$ denotes the magnetic flux quantum, while $\Phi_{x}^{\text{qfp}}$ corresponds to the external flux of QFP's small compound Josephson-junction (CJJ) loop. $E_J=\Phi_0 I_p^{\text{qfp}}/2\pi$ represents Josephson energy, where $I_p^{\text{qfp}}$ is the persistent current of the QFP.
The QFP's small CJJ loop governs the energy barrier of the potential, while the large main loop, where the persistent current is generated, controls the symmetry of the potential \cite{berkley2010scalable, harris2009compound}.
By slowly and adiabatically raising the magnetic flux in the small CJJ loop ($\Phi_{x}^{\text{qfp}}$), the barrier of potential energy is raised, and its potential energy gradually transitions from monostable to bistable. This process, known as annealing, latches the final state of the flux qubit into the corresponding QFP. In the monostable state, the QFP functions as a resonator, whereas in the bistable state, it operates as a qubit.

Neglecting constant terms, the Hamiltonian can be reformulated in terms of standard dimensionless parameters \cite{kafri2017tunable} as
\begin{equation*}
    \hat{H}_{\text{qfp}}=E_L\left(4\xi^2\frac{\hat{q}^2}{2}+\frac{\hat{\varphi}^2}{2}+\beta_L\left(\Phi_{x}^{\text{qfp}}\right)\cos{\hat{\varphi}}-\lambda\hat{\varphi}\hat{\sigma}_z\right),
\end{equation*}
where $E_L = \phi_0^2/L$ is the inductive energy, $\phi_0 = \Phi_0/(2\pi)$ is the reduced flux quantum, and $\xi = e/\phi_0 \sqrt{L/C}$ is the characteristic impedance. The operators are defined as $\hat{q} = \hat{Q}/(2e)$ and $\hat{\varphi} = \pi + \hat{\Phi}/\phi_0$. Similarly, $\varphi_{z}^{\text{qfp}} = \pi + \Phi_{z}^{\text{qfp}}/\phi_0$. Note that the definitions of $\hat{\varphi}$ and $\varphi_{z}^{\text{qfp}}$, with the $\pi$ phase shift, result in a sign flip in front of $\beta_L\left(\Phi_{x}^{\text{qfp}}\right)\cos{\hat{\varphi}}$. The coupling term is given by $\lambda = M_q\abs{I_p^q}/\phi_0$ and the screening parameter $\beta_L\left(\Phi_{x}^{\text{qfp}}\right)$ is defined as
\begin{equation*}
    \beta_L\left(\Phi_{x}^{\text{qfp}}\right) = \frac{E_J}{E_L}\cos\left(\frac{\pi\Phi_{x}^{\text{qfp}}}{\Phi_0}\right)=\frac{I_p^{\text{qfp}}L}{\phi_0}\cos\left(\frac{\pi\Phi_{x}^{\text{qfp}}}{\Phi_0}\right).
\end{equation*}

The dimensionless potential $U(\varphi)$, given by
\begin{equation*}
    U(\varphi) = \frac{\varphi^2}{2}+\beta_L\left(\Phi_{x}^{\text{qfp}}\right)\cos{\varphi}-\lambda\varphi\hat{\sigma}_z,
\end{equation*}
can be expanded around its minimum $\pm \varphi_p$, which satisfies $\frac{\partial U}{\partial \varphi}\big|_{\varphi=\pm \varphi_p}=0$, where $\varphi_p>0$ denotes the absolute value of the position of the potential minimum. Defining \(\delta\hat\varphi= \hat\varphi\mp\varphi_p=1/\sqrt{2m\Omega}(\hat a^\dagger+\hat a)\) and \(\hat q=i \sqrt{m\Omega/2}(\hat a^\dagger-\hat a)\) with annihilation operator $\hat{a}$ and creation operator $\hat{a}^\dagger$, and approximating the potential around its minimum \cite{schondorf2020flux}, the Hamiltonian becomes
\begin{equation*}
    \hat{H}_{\text{qfp}}/E_L = \Omega \hat{a}^\dagger \hat{a} + \Omega \sqrt{\frac{m\Omega}{2}} \varphi_p \left(\hat{a}^\dagger + \hat{a}\right) \hat{\sigma}_z,
\end{equation*}
where $\Omega = 2\xi \sqrt{1-\beta_L\left(\Phi_{x}^{\text{qfp}}\right)\cos(\varphi_p)}$ and $m = 1/\left(2\xi\right)^2$.

\subsection{Dispersive readout}\label{subsec:dispe_reado}

After QFP annealing, the flux qubit signal is entangled with the corresponding displaced ground states or pointer states of the QFP \cite{schondorf2020flux}, allowing the total system to be described by an efficient flux qubit Hamiltonian. More details can be found in Sec.~\ref{sec:analy}. The flux qubit signal is then read out by coupling the QFP to a resonator, with their interaction modeled as dipole coupling, resulting in a fully off-diagonal interaction, described by the Rabi Hamiltonian
\begin{equation}
    \hat{H}_{\text{Rabi}}=\frac{\omega_q}{2} \hat{\sigma}_z+\omega_r \left(\hat{a}^{\dagger}\hat{a}+\frac{1}{2}\right)+g \left(\hat{a}+\hat{a}^{\dagger}\right)\hat{\sigma}_x,
    \label{equ:H_rabi}
\end{equation}
where $\hbar$ is set to 1, $g$ is the coupling strength between the QFP and resonator, and $\omega_q$ and $\omega_r$ are the frequencies of QFP and resonator, respectively. The Pauli X matrix can be written as $\hat{\sigma}_x=\hat{\sigma}_{+}+\hat{\sigma}_{-}$, where the raising and lowering operators are defined as $\hat{\sigma}_{\pm}=\frac{1}{2}(\hat{\sigma}_{x}\pm i\hat{\sigma}_{y})$ with $\hat{\sigma}_{y}$ being the Pauli Y matrix.

Applying a rotating-frame transformation and the rotating wave approximation (RWA) reduces Eq.~\eqref{equ:H_rabi} to the Jaynes-Cummings Hamiltonian \cite{jaynes1963comparison}
\begin{equation}
    \hat{H}_{\text{JC}}=\frac{\omega_q}{2} \hat{\sigma}_z+\omega_r \left(\hat{a}^{\dagger}\hat{a}+\frac{1}{2}\right)+g \left(\hat{\sigma}_{+}\hat{a}+ \hat{\sigma}_{-}\hat{a}^{\dagger}\right).
    \label{equ:jaynes-cummings}
\end{equation}
Diagonalizing it defines the dressed eigenstates of the coupled QFP-resonator system. Bare states refer to the uncoupled product basis $\{|e,n\rangle, |g,n+1\rangle\}$, where $|g\rangle$ and $|e\rangle$ denote QFP ground and excited state, respectively, and $n$ is the resonator photon number. Dressed states are the hybridized eigenstates $\{|\pm,n\rangle\}$ that mix these bare states with detuning-controlled mixing angle $\theta_n$ defined by $\tan(\theta_n)=2g \sqrt{n+1}/\delta$, where $\delta=\omega_q-\omega_r$ is the QFP-resonator detuning (see Appendix~\ref{app:jc_dressed} for details).

Under dispersive conditions, where $\omega_r \neq \omega_q$ and $g \ll \abs{\delta}$, resonant photon absorption or emission is absent. Within this regime, we can approximate the Jaynes-Cummings Hamiltonian as \cite{blais2004cavity, gu2017microwave}
\begin{align}
	\hat{H}_{\text{disp}}=\left(\omega_r+\chi\hat{\sigma}_z\right)\left(\hat{a}^\dagger\hat{a}+\frac{1}{2}\right)+\frac{\omega_q}{2}\hat{\sigma}_z,
\end{align}
where $\chi = g^2/\delta$ is the dispersive shift.
Notably, the first term reveals that the effective resonator frequency shifts depending on the QFP state. Consequently, it is possible to perform a QND measurement of the QFP by monitoring microwave transmission near the resonator frequency.

\subsection{Measurement principle \label{subsec:measu}}

We employ the single-qubit dispersive readout methodology in the presence of two-qubit coupling, as outlined in Ref.~\cite{pommerening2020measured}, to investigate how measurement speed and power influence the effective measurement basis and its deviation from the conventional computational basis.
The measurement of qubit states proceeds as follows \cite{pommerening2020measured}. First, the resonator is initialized in a coherent state $\ket{\alpha}$ with $\alpha$ being real and positive. Subsequently, the resonator interacts with one of the qubits for a time interval $t_d = \pi/2\abs{\chi}$, which is described by the time-evolution operator $U(t) = e^{-iHt}$. Afterward, the resonator is read out via a positive operator-valued measure (POVM) with elements
\begin{eqnarray}
    & & E_{\pm}=\frac{1}{\pi} \int_{\Omega_{\pm}}d^{2}\beta \ket{\beta}\bra{\beta},
\end{eqnarray}
where $E_{+}+E_{-}=\mathbb{1}$, and integrals are over coherent states in lower $(\Omega_{\sgn(\chi)})$ and upper half-plane $(\Omega_{-\sgn(\chi)})$. Finally, the resonator is measured.

The associated superoperator describing the measurement outcome on an initial two-qubit state $\rho$ is given by
\begin{equation}
    \mathcal{E}_{\pm}\left(\rho\right)=\Tr_{\text{res}} \left[(\mathbb{1} \otimes E_{\pm}) U\left(t_d\right) (\rho \otimes \ket{\alpha}\bra{\alpha})U^{\dagger}\left(t_d\right)\right].
\label{equ:E_x}
\end{equation}
Expanding this in terms of resonator Fock states $\ket{n}$ and $\ket{m}$, we have \cite{pommerening2020measured}
\begin{equation}
    \mathcal{E}_{\pm}\left(\rho\right)=\sum_{n,m=0}^{\infty}g_{\pm}\left(m,n\right)\bra{n} U\left(t_d\right)\ket{n}\rho \bra{m} U^{\dagger}\left(t_d\right)\ket{m},
    \label{equ:E_x_expan}
\end{equation}
where
\begin{equation*}
\resizebox{\hsize}{!}{$
    g_{\pm}\left(m,n\right)
    =\mathrm{e}^{-\alpha^2} \left(\frac{\alpha^{2n}}{2n!}\delta_{mn}\mp\frac{i}{\pi}\frac{ \alpha^{n+m} \Gamma\left(\frac{m+n}{2}+1\right)}{ m! n! \left(m-n\right)} \text{odd}\left(m-n\right)\right)$}
\end{equation*}
with
\begin{equation*}
    \text{odd}\left(n\right) = \begin{cases}
	1, & \mbox{$n$ is odd}\\
	0, & \mbox{otherwise}.\end{cases}
\end{equation*}
In practical scenarios, the summation over $n$ and $m$ is truncated at a finite maximum value $N$ to ensure computational feasibility while neglecting the insignificant contributions of higher Fock states.

\section{Analysis of Single Flux Qubit Measurement \label{sec:analy}}

\begin{figure}[tb]
    \centering
    \includegraphics[width=0.93\columnwidth]{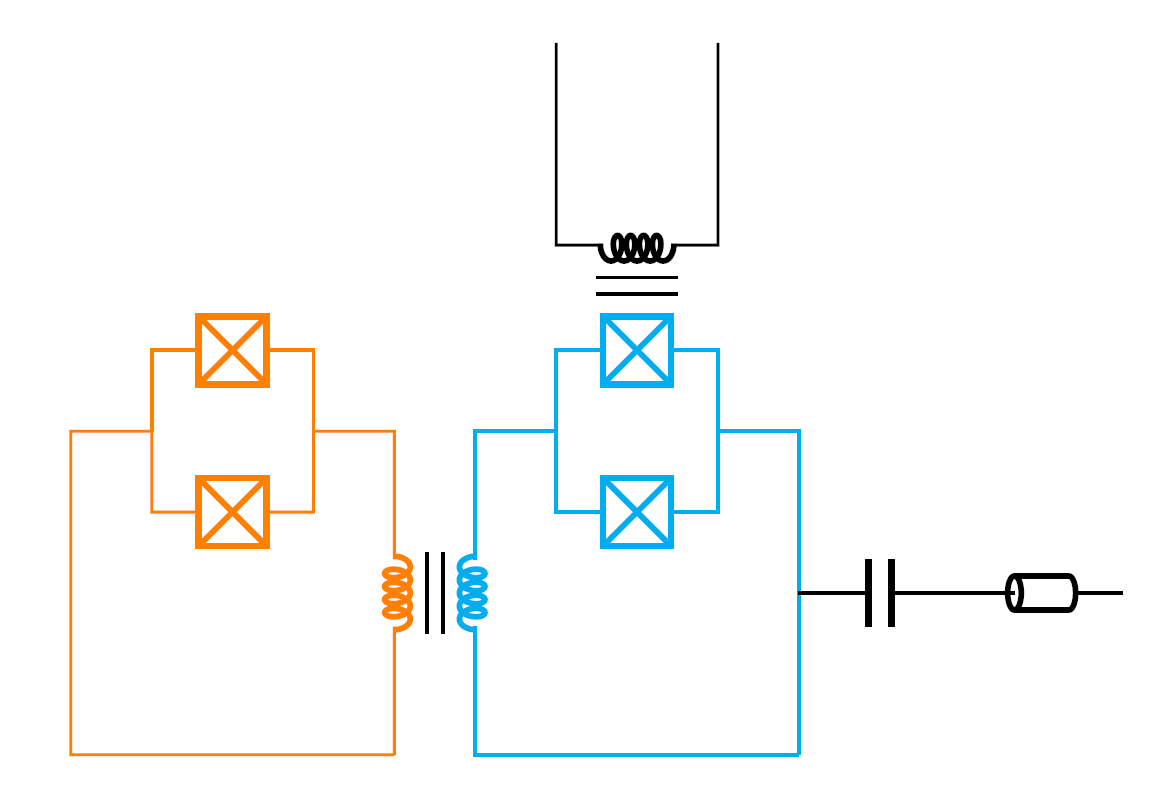}
    \put(-204,0){Flux qubit}
    \put(-114,0){QFP}
    \put(-44,18){Resonator}
    \caption{Quantum flux parametron (QFP)-mediated scheme for measuring a single flux qubit. The flux qubit is coupled to a QFP. After QFP annealing, the qubit state is latched into the QFP, which is then read out using a resonator.}
    \label{fig:modell1}
\end{figure}

This section explores the measurement process for a single flux qubit mediated by a QFP. Figure~\ref{fig:modell1} provides a schematic overview of the approach. We first examine the coupled system of flux qubit and QFP through numerical simulation to evaluate the fidelity in the flux and energy bases. Then, we assume that the adiabatic QFP annealing is successful and the flux qubit state is latched into the QFP. The coupled system is thus described by an effective flux qubit Hamiltonian. Finally, we use a resonator to read out the qubit state latched in the QFP.

\subsection{QFP annealing}

Building on the analysis in Sec.~\ref{subsec:qfp}, the total Hamiltonian describing the coupled system of flux qubit and QFP is given by
\begin{equation}
    \hat{H} =  \tilde{\Omega} \hat{a}^\dagger \hat{a} + \tilde{g}\varphi_p \left(\hat{a}^\dagger+\hat{a}\right) \hat{\sigma}_z -\frac{1}{2}\left(\epsilon_q \hat{\sigma}_z + \Delta_q \hat{\sigma}_x\right),
    \label{equ:hamiltonian_qubit_qfp}
\end{equation}
where $\epsilon_q$ and $\Delta_q$ are the energy spacing and tunneling energy of the flux qubit, respectively. The effective frequency $\tilde{\Omega}$ is defined as
\begin{equation*}
    \tilde{\Omega}=E_L\Omega=2\xi\phi_0^2/L\sqrt{1-\beta_L\left(\Phi_{x}^{\text{qfp}}\right)\cos(\varphi_p)}.
\end{equation*}
The corresponding effective coupling strength $\tilde{g}$ is then determined by $\tilde{g}=\tilde{\Omega}\sqrt{m\tilde{\Omega}/(2E_L)}$. If the two junctions in the QFP are identical, the total critical current of the QFP is $I_q^{\text{qfp}}=2I_c^{\text{qfp}}$, where $I_c^{\text{qfp}}$ is the critical current of each junction in the small CJJ loop.
Then we have
\begin{equation*}
    \beta_L\left(\Phi_{x}^{\text{qfp}}\right)=\frac{4\pi LI_c^{\text{qfp}}}{\Phi_0}\cos\left(\frac{\pi\Phi_{x}^{\text{qfp}}}{\Phi_0}\right).
\end{equation*}

Through adiabatic QFP annealing, the flux qubit and QFP evolve into an entangled state \cite{wilhelm2003asymptotic, schondorf2020flux}
\begin{equation*}
    \left(c\ket{0}+d\ket{1}\right)\otimes \ket{0}\longrightarrow\left(c_{\text{eff}}\ket{0,L}+d_{\text{eff}}\ket{1,R}\right),
\end{equation*}
where the flux qubit state becomes entangled with the pointer states of the QFP. The system is then described by an effective flux qubit Hamiltonian (further details are provided in Appendix~\ref{subsec:effec_hamil_coupl_flux_qfp}).
Under the Gaussian approximation, the fidelity measured in the flux basis for a measurement time interval $T$ is given by \cite{schondorf2020flux}
\begin{equation}
    \mathcal{F}_T=\Phi\left(\frac{\varphi_p(T)}{{\sigma}(T)}\right),
\end{equation}
where ${\sigma}(T)=\left[\sqrt{1-\beta_L(T)\cos(\varphi_p(T))}/\xi\right]^{-1/2}$ is the standard deviation of Gaussian wave packet, $\varphi_p(T)$ is the positive position of double-well potential minimum at time $T$, and $\Phi(x) = \frac{1}{\sqrt{2\pi}}\int_{-\infty}^{x}e^{-t^2/2} dt$ is the normal cumulative distribution function.

In the qubit energy eigenbasis, the positive position of potential minimum is given by $\tilde{\varphi}_p(T) = \left[\cos(\theta_{q})-i\sin(\theta_{q})\right]\varphi_p(T)$, where $\theta_{q}$ is defined as $\tan (\theta_{q}) =\Delta_{q}/\epsilon_{q}$, and the fidelity becomes
\begin{equation}
    \tilde{\mathcal{F}}_T=\Phi\left(\frac{\tilde{\varphi}_p(T)}{\tilde{{\sigma}}(T)}\right),
    \label{eq:annealing_fid}
\end{equation}
where $\tilde{{\sigma}}(T)=\left[\sqrt{1-\beta_L(T)\cos(\tilde{\varphi}_p(T))}/\xi\right]^{-1/2}$.

\begin{figure}[tb]
    \centering
    \includegraphics[width=0.93\columnwidth]{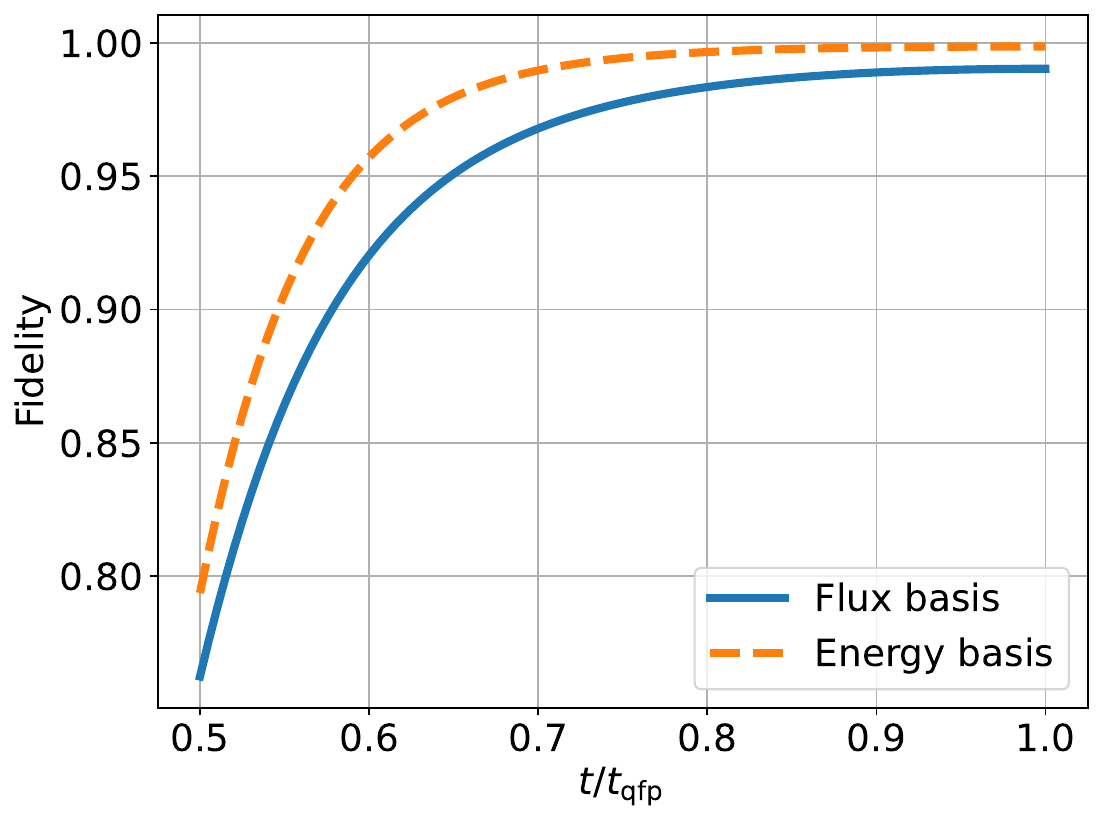}
    \put(-0.93\columnwidth, 165){\textbf{(a)}}\\
    \includegraphics[width=0.93\columnwidth]{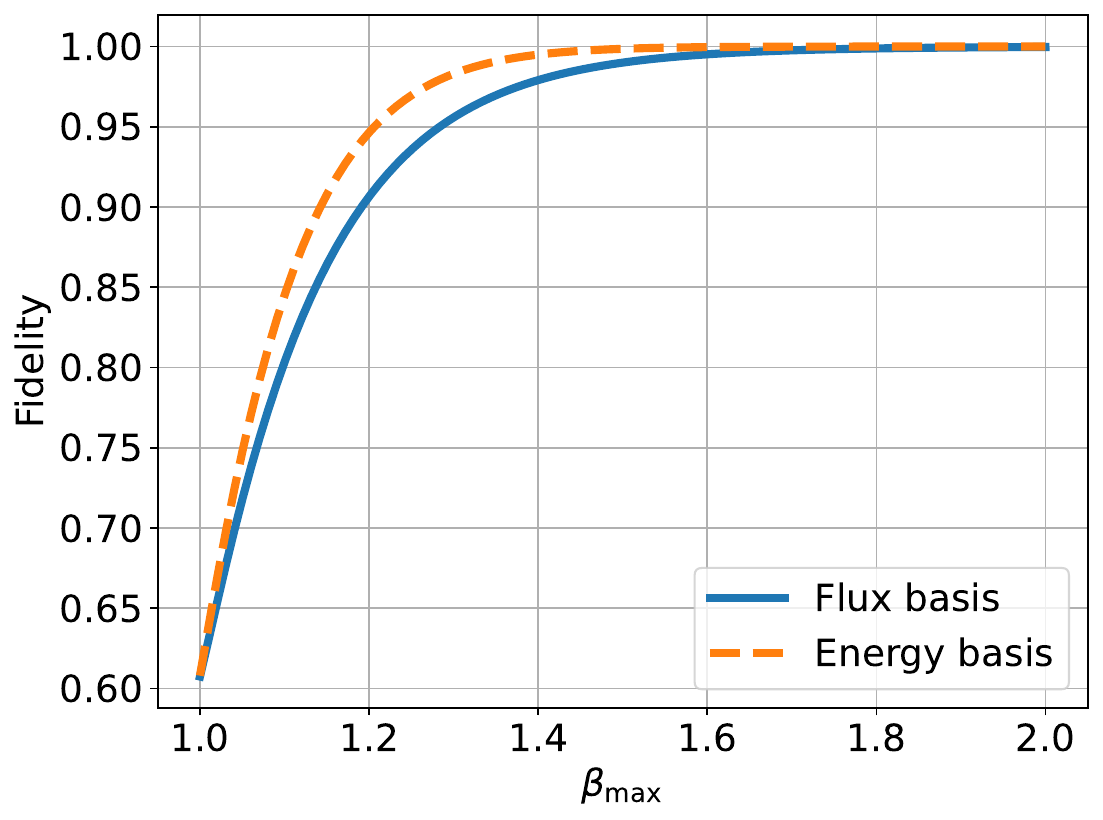}
    \put(-0.93\columnwidth, 165){\textbf{(b)}}
    \caption{QFP annealing fidelity as a function of (a) normalized annealing time $t/t_{\mathrm{qfp}}$ for $\beta_{\mathrm{max}}=1.5$ and (b) maximal screening parameter $\beta_{\mathrm{max}}$ at the total annealing time $t_{\mathrm{qfp}}$. Fidelity is evaluated in both flux and energy bases for $\Delta_{q}/\epsilon_{q}=4$ and $\xi=0.4$.}
    \label{fig:an_t_betamax}
\end{figure}

Figure~\ref{fig:an_t_betamax} illustrates the fidelity evolution as a function of $t/t_{\text{qfp}}$ and $\beta_{\text{max}}$, where $t_{\text{qfp}}$ is the total annealing time. As shown in Fig.~\ref{fig:an_t_betamax}(a), the fidelity consistently increases over time in both flux and energy bases, saturating near the completion of QFP annealing process. At the full annealing time $t_{\text{qfp}}$, the flux qubit state is latched into the QFP ground state, achieving fidelities approaching unity.
Moreover, the energy basis outperforms the flux basis by achieving higher fidelity.
As illustrated in Fig.~\ref{fig:an_t_betamax}(b), the fidelity in the energy basis reaches its maximum at $\beta_{\text{max}} = 1.4$, while in the flux basis it peaks at a slightly higher value, $\beta_{\text{max}} = 1.6$. For $\beta_{\text{max}}$ values below these optima, the potential barrier during QFP annealing is insufficient, resulting in less distinguishable states and subsequently lower fidelities.
These results demonstrate that adiabatic QFP annealing can successfully latch the flux qubit state into the QFP in both the flux and energy bases, with higher efficiency observed in the energy basis.

\subsection{Measurement in different bases \label{subsec:measu_in_diffe}}

After annealing, the resulting coupled system of flux qubit and QFP is described by an effective flux qubit Hamiltonian \cite{wilhelm2003asymptotic} (see Appendix~\ref{subsec:effec_hamil_coupl_flux_qfp} for further details)
\begin{equation}
    \hat{H}_{\text{eff}}=-\frac{1}{2}\left(\epsilon_q\hat{\sigma}_z+\Delta_{\text{eff}}\hat{\sigma}_x\right)
    \label{eq:effec_flux_qubit_hamil}
\end{equation}
where $\Delta_{\text{eff}}$ is the effective tunneling amplitude reduced compared to the original qubit tunneling $\Delta_q$. This reduction is characterized by a Franck-Condon-type factor of $e^{-\eta}$ ($\Delta_{\text{eff}} = \Delta_{q}e^{-\eta}$), where $\eta$ is determined by the specific QFP design \cite{wilhelm2003asymptotic}.

The QFP is then read out via a coupled resonator, as demonstrated in Fig.~\ref{fig:modell1}. The Hamiltonian of the system is described by 
\begin{equation}
    \hat{H}=-\frac{1}{2}\left(\epsilon_q\hat{\sigma}_z+\Delta_{\text{eff}}\hat{\sigma}_x\right)+\omega_r \hat{a}^\dagger\hat{a}+ g \hat{\sigma}_z  \left(\hat{a} +\hat{a}^\dagger\right),
    \label{eq:H_singleq_qfp}
\end{equation}
where $\omega_r$ is the resonator frequency and $g$ is the coupling strength.
Transforming it into the qubit energy eigenbasis and applying the RWA together with a displacement transformation yields (see Appendix~\ref{app:single_qubit_disp_steps}):
\begin{equation}
    \tilde{\hat{H}}_{(e)}=-\left[\frac{\delta}{2}+\chi\left(\hat{a}^{\dag}\hat{a}+\frac{1}{2}\right)\right]\tilde{\hat{\sigma}}_z,
    \label{equ:H_3}
\end{equation}
where $\chi = g^2\sin^2(\theta_{\text{eff}})/\delta$, $\theta_{\mathrm{eff}}$ is defined as $\tan(\theta_{\mathrm{eff}})=\Delta_{\mathrm{eff}}/\epsilon_q$, and 
$\delta=\omega_{\mathrm{eff}}-\omega_r$ with
$\omega_{\mathrm{eff}}=\sqrt{\epsilon_q^2+\Delta_{\mathrm{eff}}^2}$.
Transforming the Hamiltonian back into the flux basis, we obtain
\begin{equation}
    \hat{H}_{(f)}=-\left[\frac{\delta}{2}+\chi\left(\hat{a}^{\dag}\hat{a}+\frac{1}{2}\right)\right]\left[\cos(\theta_{\text{eff}})\hat{\sigma}_z+\sin(\theta_{\text{eff}})\hat{\sigma}_x\right].
    \label{equ:H_3_b}
\end{equation}

In the following, we quantify how closely the QFP-mediated dispersive readout reproduces an ideal nonselective L\"uders update \cite{luders2006concerning,fiorentino2026beyond} in a chosen candidate basis $\mathcal{B}$ by evaluating a fixed-input, state-level backaction fidelity.
For a chosen probe state $\rho_{\rm in}$, the two outcomes define unnormalized conditional postmeasurement states $\tilde{\rho}_{\pm}=\mathcal{E}_{\pm}(\rho_{\rm in})$ with probabilities $p_{\pm}=\Tr\,\tilde{\rho}_{\pm}$, and the corresponding nonselective postmeasurement state is $\rho_{\rm out}=\tilde{\rho}_{+}+\tilde{\rho}_{-}$. 
As an ideal reference for a projective measurement of the target qubit $q_2$ in basis $\mathcal{B}$, we use the associated L\"uders instrument with projectors $\Pi^{(\mathcal{B})}_{0}=\mathbb{1}_{q_1}\otimes \ket{0_{\mathcal{B}}}\!\bra{0_{\mathcal{B}}}_{q_2}$ and $\Pi^{(\mathcal{B})}_{1}=\mathbb{1}_{q_1}\otimes \ket{1_{\mathcal{B}}}\!\bra{1_{\mathcal{B}}}_{q_2}$, yielding the ideal nonselective output 
\begin{equation}
    \rho^{\rm(id)}_{\rm out}=\sum_{i=0}^{1}\Pi^{(\mathcal{B})}_{i}\rho_{\rm in}\Pi^{(\mathcal{B})}_{i}.
\end{equation}
When assessing backaction on subsystem $s$, we compare the reduced nonselective outputs $\rho_{\rm red}=\Tr_{\bar s}[\rho_{\rm out}]$ and $\rho^{\rm(id)}_{\rm red}=\Tr_{\bar s}[\rho^{\rm(id)}_{\rm out}]$, where $\bar{s}$ denotes the complement of $s$, using the state fidelity
\begin{equation}
F(\rho_{\rm red},\rho^{\rm(id)}_{\rm red})
=\left(\Tr\sqrt{\sqrt{\rho_{\rm red}}\,\rho^{\rm(id)}_{\rm red}\,\sqrt{\rho_{\rm red}}}\right)^2.
\label{eq:Fmeas}
\end{equation}

This fidelity equals unity if and only if the reduced nonselective outputs coincide for the chosen probe state, and it quantifies outcome-averaged backaction, e.g., measurement-induced dephasing, on the retained subsystem.
Unless stated otherwise, we evaluate Eq.~\eqref{eq:Fmeas} using the coherence probe $\rho_{\rm in}=\ket{+_{\mathcal{B}}}\!\bra{+_{\mathcal{B}}}$ in the single-qubit case and $\rho_{\rm in}=\ket{+_{\mathcal{B}}}\!\bra{+_{\mathcal{B}}}\otimes\ket{+_{\mathcal{B}}}\!\bra{+_{\mathcal{B}}}$ in the two-qubit case, where $\ket{+_{\mathcal{B}}}=(\ket{0_{\mathcal{B}}}+\ket{1_{\mathcal{B}}})/\sqrt{2}$.
These superposition inputs are standard probes for diagnosing measurement-induced dephasing \cite{gambetta2006qubit} and its extensions to joint measurements \cite{lalumiere2010tunable}.
In the numerics, this is implemented by keeping the same matrix form $\rho_{\rm in}=\ket{+}\!\bra{+}$ (single-qubit) and $\rho_{\rm in}=\ket{+}\!\bra{+}\otimes\ket{+}\!\bra{+}$ (two-qubit) in the local basis used to express the effective Hamiltonian for the chosen $\mathcal{B}$. Equivalently, each $\mathcal{B}$ run prepares the basis-adapted physical probe $\ket{+_{\mathcal{B}}}$. Thus, the $\mathcal{B}$ dependence of Eq.~\eqref{eq:Fmeas} reflects backaction on different physical coherences and closeness to the corresponding L\"uders dephasing channel, not a passive change of representation.

\begin{figure}[tb]
    \centering
    \includegraphics[width=0.93\columnwidth]{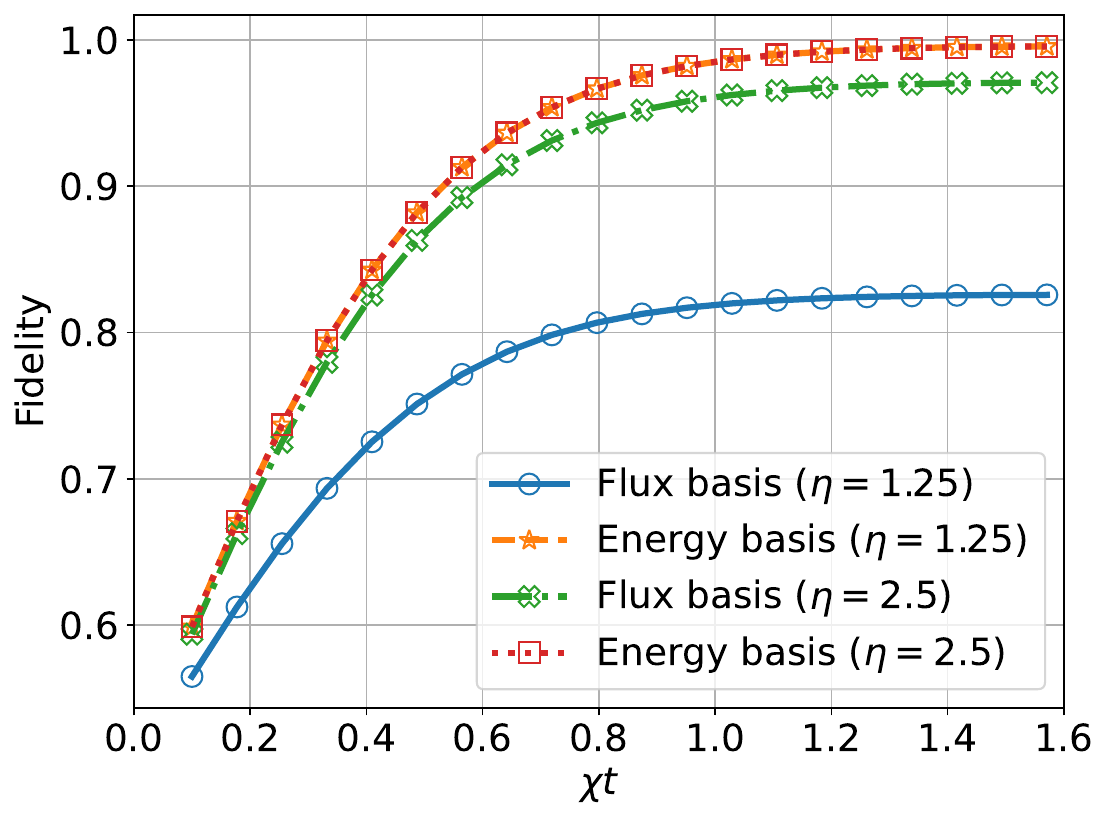}
    \put(-0.93\columnwidth, 165){\textbf{(a)}}\\
    \includegraphics[width=0.93\columnwidth]{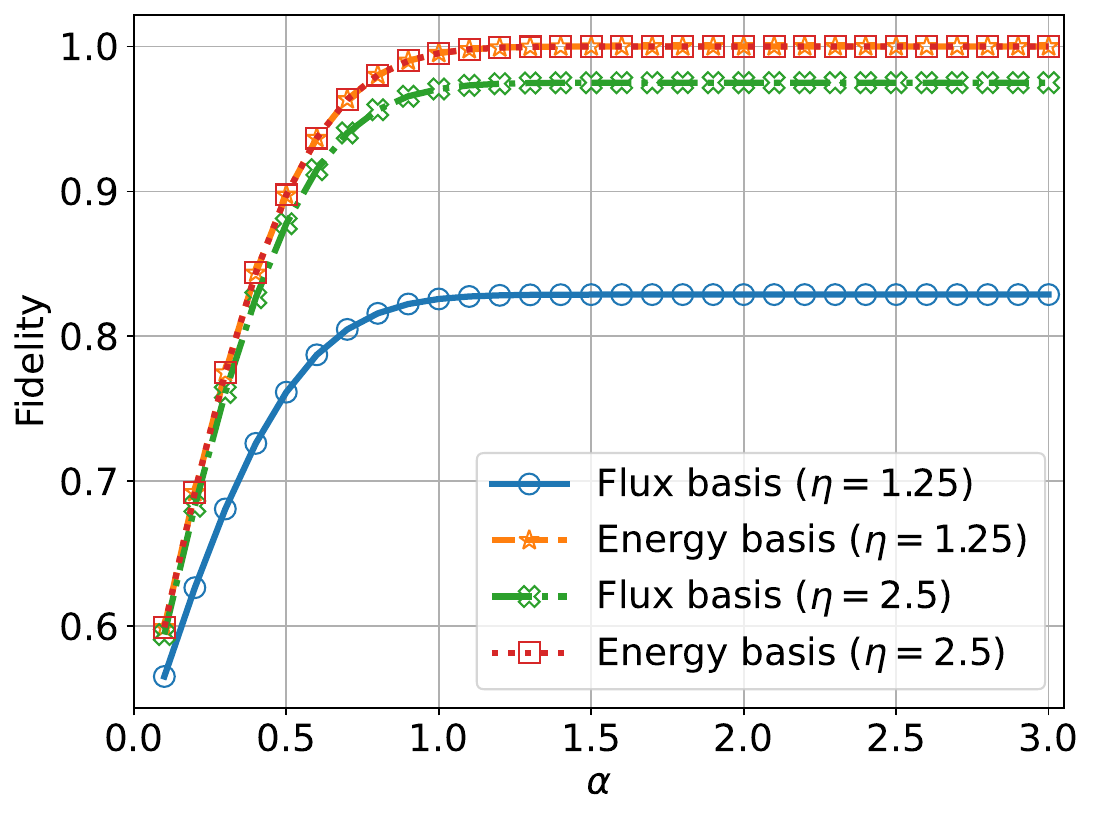}
    \put(-0.93\columnwidth, 165){\textbf{(b)}}
    \caption{Fidelity calculated using the Hamiltonian expressed in both flux and energy bases for $\eta=1.25$ and $\eta=2.5$, with $N=27$, $\Delta_{q}/\epsilon_{q}=4$, and $\delta/g=8$. Fidelity as a function of (a) $\chi t$ for $\alpha=1$ and (b) $\alpha$ at the measurement time $t = t_d$.}
    \label{fig:1q_chit_alpha_fq}
\end{figure}

Figure~\ref{fig:1q_chit_alpha_fq} presents numerical results exploring measurement fidelity as a function of $\chi t$ and $\alpha$. In Fig.~\ref{fig:1q_chit_alpha_fq}(a), we observe that increasing $\chi t$ leads to a significant improvement in fidelity, reaching a maximum value at the measurement time $t = t_d$. This indicates that, for the chosen probe state and parameters, increasing the interaction time improves agreement with the ideal Lüders backaction. Similar trends are observed for the resonator coherent state amplitude $\alpha$ in Fig.~\ref{fig:1q_chit_alpha_fq}(b), where larger values of $\alpha$ yield improved achievable fidelities, while smaller values of $\alpha$ lead to reduced fidelities regardless of the measurement basis. Crucially, energy basis consistently outperforms flux basis across all $\chi t$ and $\alpha$ values. This shows that the nonselective output is closer to an ideal energy-basis dephasing channel than to a flux-basis dephasing channel, indicating reduced non-QND mixing in the energy-basis. Furthermore, as $\eta$ increases, the difference between flux and energy basis decreases. As shown in Fig.~\ref{fig:1q_chit_alpha_fq}, larger $\eta$ enhances fidelity in the flux basis but leaves it unchanged in the energy basis. The underlying reason is that, as $\eta$ increases, $\Delta_{\text{eff}}$ approaches zero, causing the flux basis to more closely approximate the energy basis.

\section{Measurement of Coupled Flux Qubit System \label{sec:measu}}

In this section, we delve into a system of two coupled flux qubits denoted as \(q_1\) and \(q_2\). Each qubit is coupled to a QFP: specifically, \(q_1\) couples to QFP\textsubscript{1}, and \(q_2\) couples to QFP\textsubscript{2}. Our primary objective is to read out the state of a single flux qubit \(q_2\) in this coupled system. We aim to determine whether the QFP-mediated measurement ends up in a single- or two-qubit eigenbasis. Two distinct models are explored: sequential and simultaneous measurements.
In the sequential measurement model, QFP\textsubscript{2} undergoes adiabatic annealing, while QFP\textsubscript{1} remains unaffected. After adiabatic QFP\textsubscript{2} annealing, the measurement of \(q_2\) is performed using a resonator.
In contrast, both QFP\textsubscript{1} and QFP\textsubscript{2} undergo adiabatic annealing simultaneously in the simultaneous measurement model. We conduct theoretical analysis and numerical simulations to assess the fidelity of \(q_2\) under both measurement conditions.

\subsection{Sequential measurement}

\begin{figure}[tb]
    \centering
    \includegraphics[width=0.99\columnwidth]{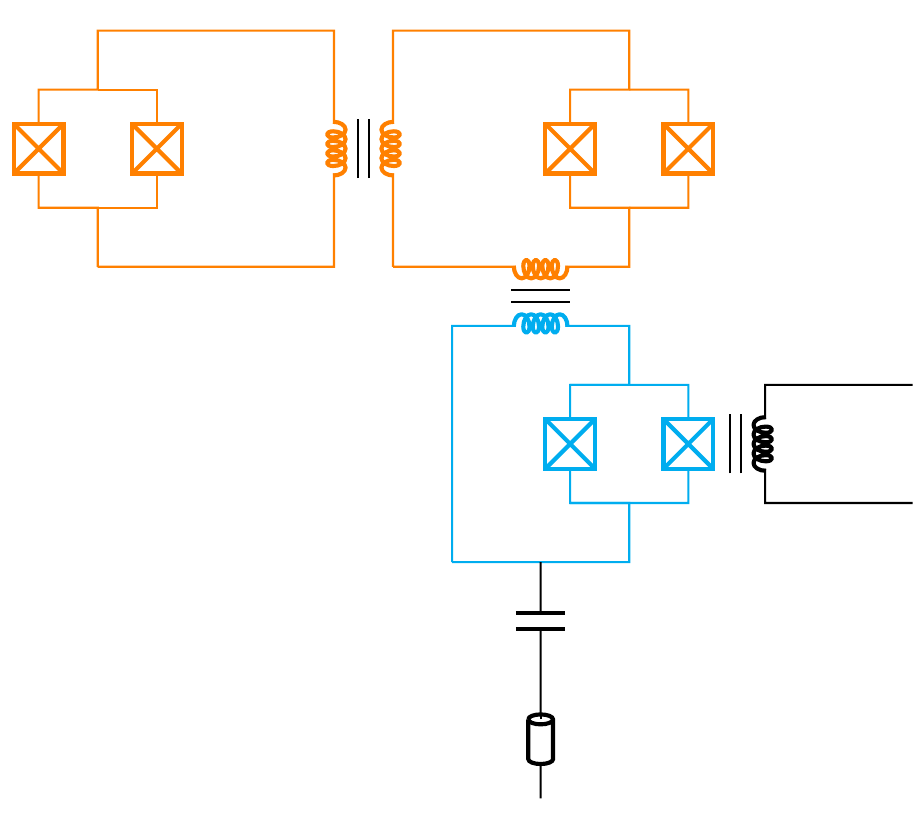}
    \put(-215,212){Flux qubit $q_1$}
    \put(-136,212){Flux qubit $q_2$}
    \put(-88,48){QFP\textsubscript{2}}
    \put(-93,15){Resonator}
    \caption{Sequential measurement model. Two flux qubits $q_1$ and $q_2$ are coupled. QFP\textsubscript{2} undergoes adiabatic annealing to latch the state of $q_2$. Subsequently, the state of QFP\textsubscript{2} is read out using a resonator.}
    \label{fig:seq_measu}
\end{figure}
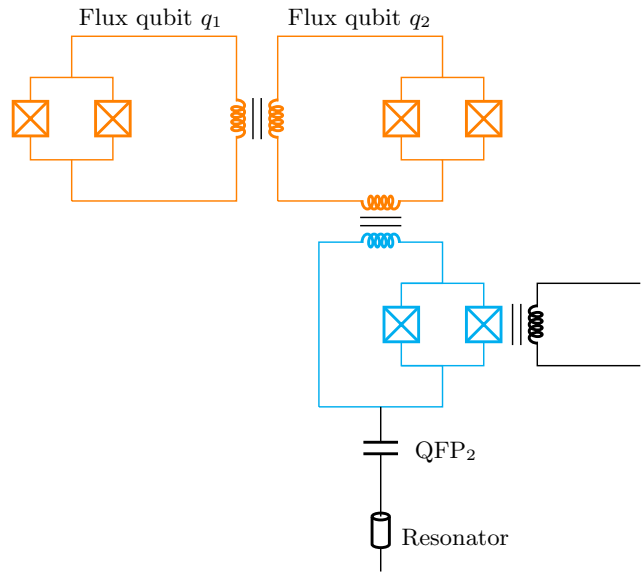

The sequential measurement model is presented in Fig.~\ref{fig:seq_measu}. Before introducing the Hamiltonian, it is important to clarify the underlying assumptions. In particular, we consider a scenario where the first qubit, $q_1$, remains essentially static or strongly biased, rendering its intrinsic dynamics negligible over the relevant timescales. Under these conditions, $q_1$ can be treated as a quasi-classical parameter, allowing its Hamiltonian to be omitted without compromising the essential physics. Retaining only the qubit-qubit interaction ensures that the influence of the first qubit on the second remains accurately captured, thereby yielding a simplified yet physically meaningful description of the system. Furthermore, we assume that adiabatic QFP\textsubscript{2} annealing has been successfully performed, with the state of $q_2$ fully latched into QFP\textsubscript{2}. The focus then shifts to the readout process of QFP\textsubscript{2} using a resonator, accompanied by an analysis of fidelity across various bases. The total Hamiltonian in the flux basis is given by
\begin{eqnarray}
    \hat{H}&=&-\frac{1}{2}\left(\epsilon_{2} \hat{\sigma}_2^z+\Delta_{\text{eff,2}}\hat{\sigma}_2^x\right)+J\hat{\sigma}_1^z\hat{\sigma}_2^z \nonumber+g\hat{\sigma}_2^z\left(\hat{a}^{\dag}+\hat{a}\right)\\
    & & +\omega_r \hat{a}^{\dag}\hat{a},
    \label{eq:hamil_seque_measu}
\end{eqnarray}
where $J$ is the coupling strength between two flux qubits, $g$ represents coupling strength between $q_2$ and resonator, and $\omega_r$ is resonator frequency. In Appendix~\ref{subsec:backa_qfp_flux_qub}, we demonstrate that $J$ is not renormalized.

Transforming Eq.~\eqref{eq:hamil_seque_measu} to the $q_2$ energy basis in the rotating frame and applying the standard displacement and dispersive unitary transformations yields (see Appendix~\ref{app:seq_q2_dress} for details):
\begin{equation}
    \tilde{\hat{H}}_{(e_2)}\approx-\frac{\hat{\delta}_{\text{eff2,n}}}{2}\tilde{\hat{\sigma}}_2^z+J\hat{\sigma}_1^z\left[\cos(\theta_{\text{eff,2}}) \tilde{\hat{\sigma}}_2^z-\sin(\theta_{\text{eff,2}}) \tilde{\hat{\sigma}}_2^x\right],
    \label{equ:Ht2_q2eig}
\end{equation}
where $\hat{\delta}_{\text{eff2,n}} = \delta_{\text{eff,2}}+\chi\left(2\hat{n}+1\right)$ with $\delta_{\mathrm{eff},2}=\omega_{\mathrm{eff},2}-\omega_r$, $\omega_{\mathrm{eff},2}=\sqrt{\epsilon_2^2+\Delta_{\mathrm{eff},2}^2}$, the resonator photon-number operator $\hat{n} =\hat{a}^{\dag}\hat{a}$, and the effective dispersive shift $\chi=g^2\sin^2(\theta_{\text{eff,2}})/ \delta_{\text{eff,2}}$. Here, the effective mixing angle satisfies $\tan(\theta_{\mathrm{eff},2})=\Delta_{\mathrm{eff},2}/\epsilon_2$.

To study the fidelity in different bases, we express the Hamiltonian (Eq.~\eqref{equ:Ht2_q2eig}) in the flux basis as
\begin{eqnarray}
    \hat{H}_{(f)}&\approx&-\frac{\hat{\delta}_{\text{eff2,n}}}{2}\left[\cos(\theta_{\text{eff,2}}) \hat{\sigma}_2^z+\sin(\theta_{\text{eff,2}}) \hat{\sigma}_2^x\right]\nonumber\\
    & &+J\hat{\sigma}_1^z\hat{\sigma}_2^z.
    \label{equ:Hf}
\end{eqnarray}
Similarly, the Hamiltonian can be further written in the $q_1$ energy basis (or $q_1$-$q_2$ energy basis)
\begin{eqnarray}
    \tilde{\hat{H}}_{(e)}&\approx&-\frac{\hat{\delta}_{\text{eff2,n}}}{2}\tilde{\hat{\sigma}}_2^z+J\left[\cos(\theta_1) \tilde{\hat{\sigma}}_1^z-\sin(\theta_1) \tilde{\hat{\sigma}}_1^x\right]\nonumber\\
    & &\otimes \left[\cos(\theta_{\text{eff,2}}) \tilde{\hat{\sigma}}_2^z-\sin(\theta_{\text{eff,2}}) \tilde{\hat{\sigma}}_2^x\right],
    \label{equ:ht3_q1q2}
\end{eqnarray}
where $\theta_1$ is defined as $\tan(\theta_1)=\Delta_1/\epsilon_1$ with energy spacing $\epsilon_1$ and tunneling energy $\Delta_{1}$ of $q_1$.
Then we can investigate the fidelity in different bases with the measurement operator
\begin{equation*}
    \mathcal{E}_{\pm}(\rho)=\Tr_{\text{res}}\Tr_{\text{$q_1$}} \left[(\mathbb{1} \otimes E_{\pm})U(t_d)(\rho\otimes\ket{\alpha} \bra{\alpha})\hat{U}^\dagger(t_d)\right].
    \label{equ:sup_2q}
\end{equation*}
Compared to the previous case, we need to additionally trace out $q_1$ since we are only interested in the fidelity of $q_2$.

\begin{figure}[tb]
    \centering
    \includegraphics[width=0.92\columnwidth]{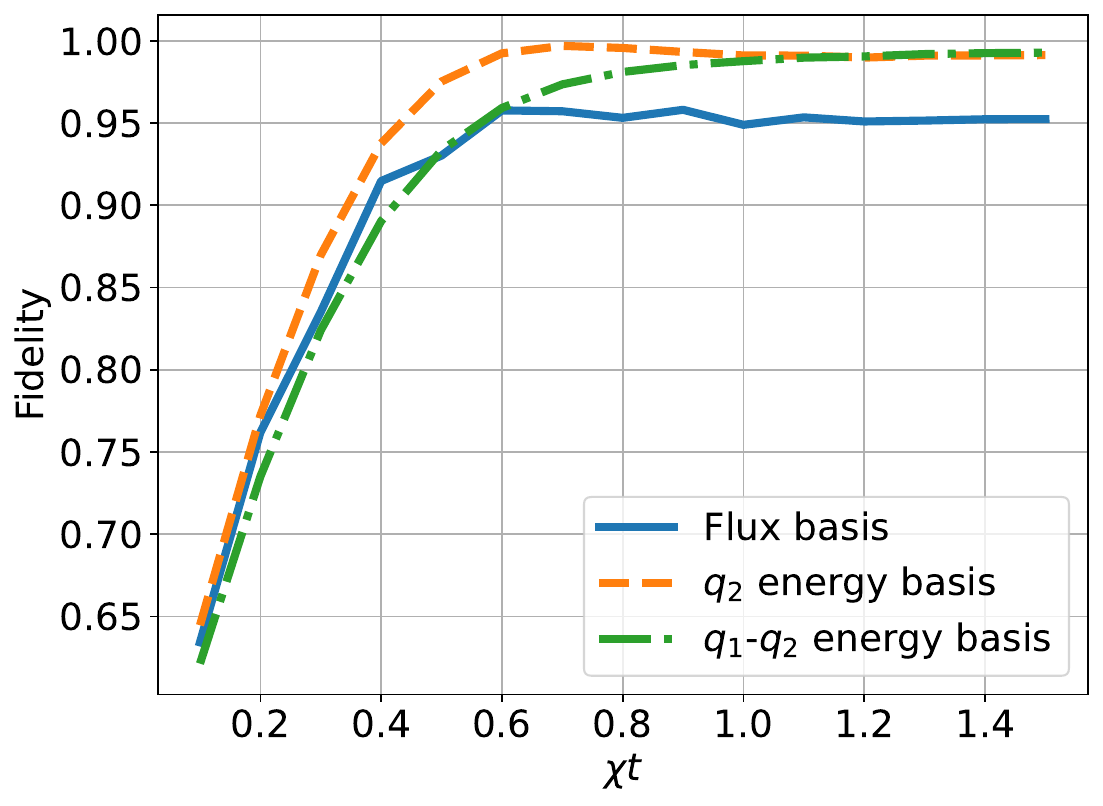}
    \put(-0.93\columnwidth, 160){\textbf{(a)}}\\
    \includegraphics[width=0.92\columnwidth]{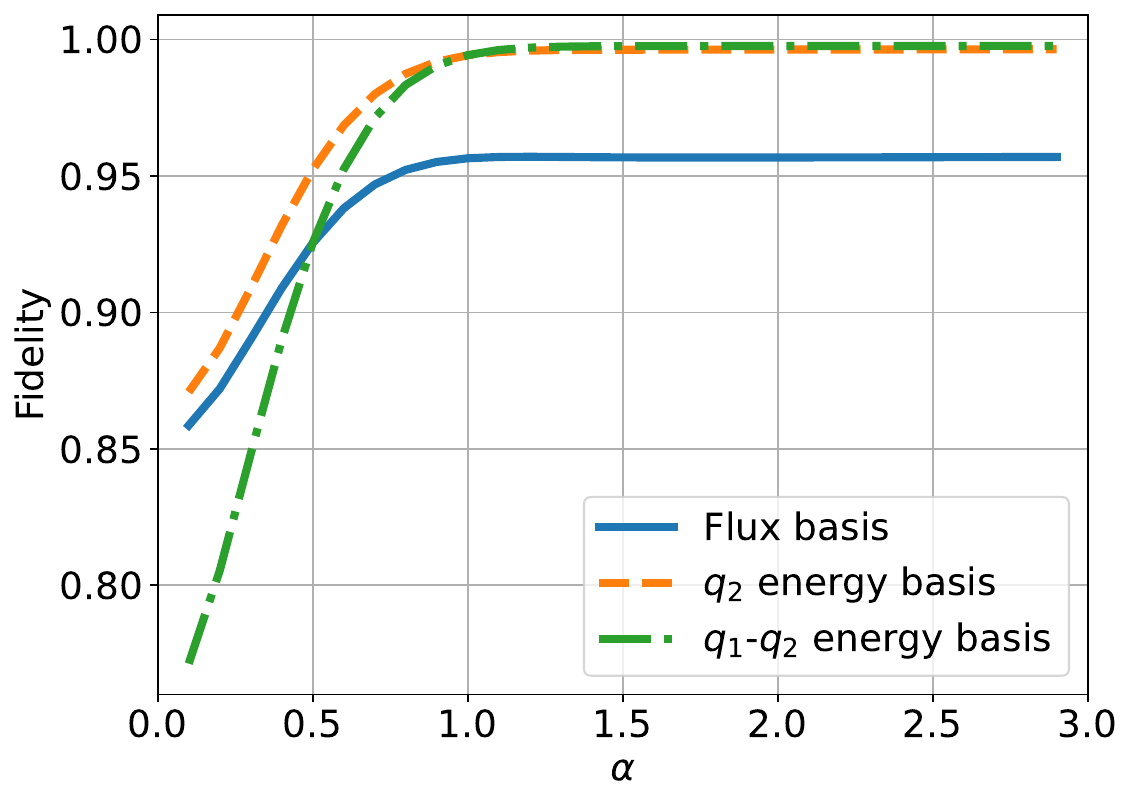}
    \put(-0.93\columnwidth, 160){\textbf{(b)}}\\
    \includegraphics[width=0.92\columnwidth]{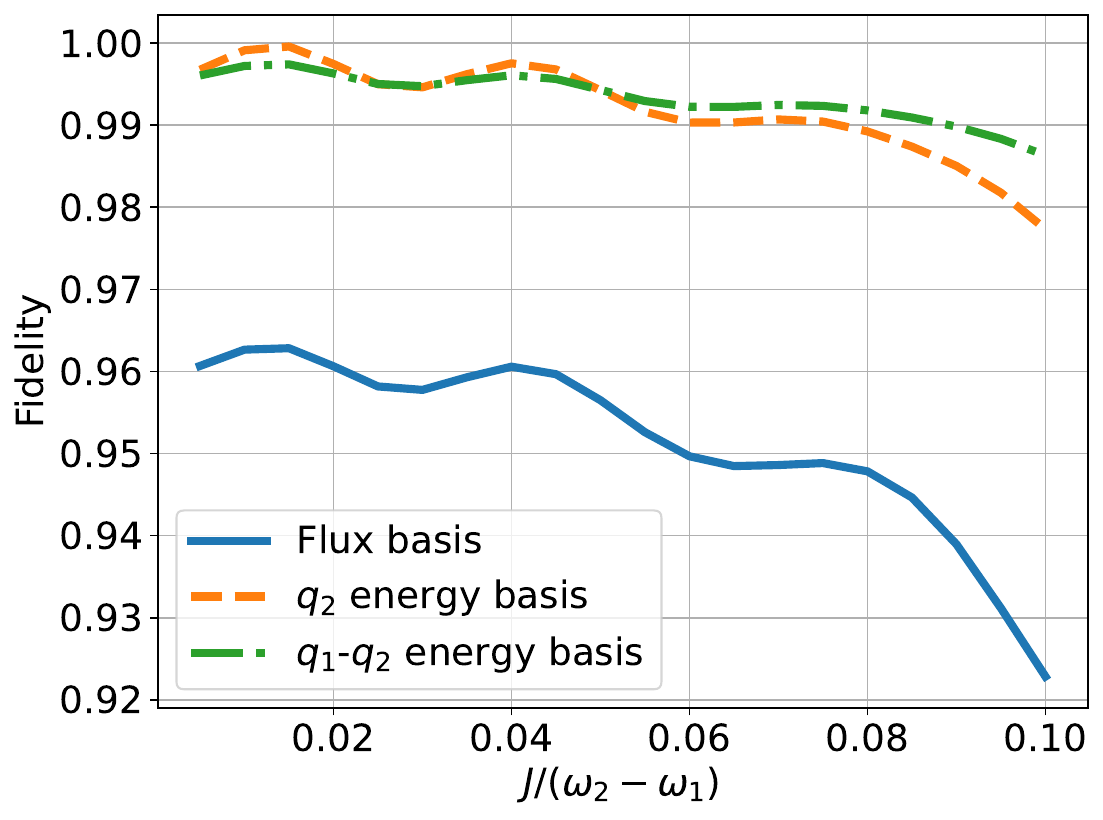}
    \put(-0.93\columnwidth, 160){\textbf{(c)}}
    \vspace{-.5em}
    \caption{Fidelity calculated using the Hamiltonian expressed in the flux basis, $q_2$ energy basis, and $q_1$-$q_2$ energy basis with $N=27$, $\eta=1.25$, $\Delta_{2}/\epsilon_2 = 1$, and $\delta/g=8$. Fidelity as a function of (a) $\chi t$ for $J/(\omega_{2}-\omega_{1})=0.05$ and $\alpha=1$; (b) $\alpha$ for $J/(\omega_{2}-\omega_{1})=0.05$ and $t = t_d$; and (c) $J/(\omega_{2}-\omega_{1})$ for $t = t_d$ and $\alpha=1$.}
    \label{fig:2qo_fq2q1q2}
\end{figure}

Figure~\ref{fig:2qo_fq2q1q2} displays the fidelity dependence on $\chi t$, $\alpha$, and $J/(\omega_{2}-\omega_{1})$, where $\omega_{1}=\sqrt{\epsilon_{1}^2+\Delta_{1}^2}$.
As shown in Fig.~\ref{fig:2qo_fq2q1q2}(a), the $q_2$ energy basis exhibits the highest overall fidelity, peaking at the measurement time $t = t_d/2$, and then displaying a slight decline.
This behavior might be attributed to the interaction between $q_1$ and $q_2$.
Due to the dressed state formation, we measure the coupled system instead of $q_2$ individually.
This distinction disappears only when the QND condition holds. Therefore, for longer measurement times, the $q_1$-$q_2$ energy basis might be more appropriate.
However, the $q_1$-$q_2$ energy basis reaches maximum fidelity more slowly than the $q_2$ energy basis. Additionally, for shorter measurement times $(t\lesssim t_d/\pi)$, the flux basis outperforms the $q_1$-$q_2$ energy basis.
At $\alpha \approx 1$, all bases achieve the maximum fidelity and remain constant, as shown in Fig.~\ref{fig:2qo_fq2q1q2}(b). However, both energy bases outperform the flux basis. A basis crossover between flux basis and $q_1$-$q_2$ energy basis occurs around $\alpha\approx0.5$, whereas the crossover between the two energy bases is less pronounced. This is potentially due to the chosen parameter values of $t = t_d$ (see Fig.~\ref{fig:2qo_fq2q1q2}(a)) and $J/(\omega_{2}-\omega_{1})=0.05$ (see Fig.~\ref{fig:2qo_fq2q1q2}(c)).
Finally, as shown in Fig.~\ref{fig:2qo_fq2q1q2}(c), we observe consistently higher fidelity in the $q_2$ and $q_1$-$q_2$ energy bases compared to the flux basis. Stronger coupling between $q_1$ and $q_2$ leads to mutual influence during measurement, reducing fidelity. The basis crossover between two energy bases occurs at $J/(\omega_{2}-\omega_{1})\approx0.05$, after which the $q_1$-$q_2$ energy basis exhibits higher fidelity.

In the following, we delve into the validity of the RWA within both bare and dressed bases using the method described in Ref.~\cite{pommerening2020measured}. Our investigation commences with the Hamiltonian operator expressed in the energy basis of the measured flux qubit (specifically, the $q_2$ energy basis, as given in Eq.~\eqref{equ:Ht2_q2eig}). We then extend our analysis to consider the Hamiltonian operator formulated in the energy basis of both flux qubits ($q_1$-$q_2$ energy basis, detailed in Eq.~\eqref{equ:ht3_q1q2}).

\subsubsection{Energy basis of the measured flux qubit}

To define the dressed basis analytically, we work within fixed photon-number manifolds and replace the photon-number operator $\hat{n}=\hat{a}^{\dag}\hat{a}$ by its eigenvalue $n\in\{0,1,2,\dots\}$ within each Fock manifold, so that $\hat{\delta}_{\text{eff2,n}}$ becomes a constant number $\delta_{\text{eff2,n}}$ within each manifold.
The full operator $\hat{n}$ is restored when evaluating the qubit-resonator dynamics and the associated measurement channel.
Diagonalizing the resulting $n$-resolved Hamiltonian Eq.~\eqref{equ:Ht2_q2eig} then yields the dressed eigenpairs that define the dressed basis.
The spectrum consists of two doublets with frequencies $\omega_{n\pm}=\sqrt{\big(\delta_{\mathrm{eff}2,n}/2\pm J_{zz}\big)^2+J_{zx}^2}$, where the effective couplings are $J_{zz}=J\cos(\theta_{\mathrm{eff},2})$ and $J_{zx}=J\sin(\theta_{\mathrm{eff},2})$. Further details are provided in Appendix~\ref{app:seq_q2_dress}.

The validity of the RWA in the dressed basis requires
\begin{equation*}
    \left\Vert\frac{J_{zx}~\chi \left(\hat{a}^{\dag}\hat{a}+\frac{1}{2}\right) }{\left(\hat{\delta}_{\text{eff2,n}}/2 \pm J_{zz}\right)\left(\delta_{\text{eff,2}}/2 \pm J_{zz}\right)+J_{zx}^2}\right\Vert \ll 1.
    \label{bed:mo2d}
\end{equation*}
Similarly, the RWA condition based on the bare basis is
\begin{equation*}
    \left\Vert\frac {J_{zx}}{\left(\hat{\delta}_{\text{eff2,n}}/2 \pm J_{zz}\right)}\right\Vert \ll 1.
    \label{bed:mo2b}
\end{equation*}
In photon-number manifold $n$, the crossover occurs when
\begin{equation*}
    \left\Vert \chi \left(\hat{a}^{\dag}\hat{a}+\frac{1}{2}\right) \right\Vert=\sqrt{\left(\delta_{\text{eff,2}}/2 \pm J_{zz}\right)^2+J_{zx}^2},
    \label{bed:fq2o_fq2}
\end{equation*}
so that measurement-induced phase separation competes with the coherent hybridization scale.

\begin{figure}[t]
    \centering
    \includegraphics[width=0.93\columnwidth]{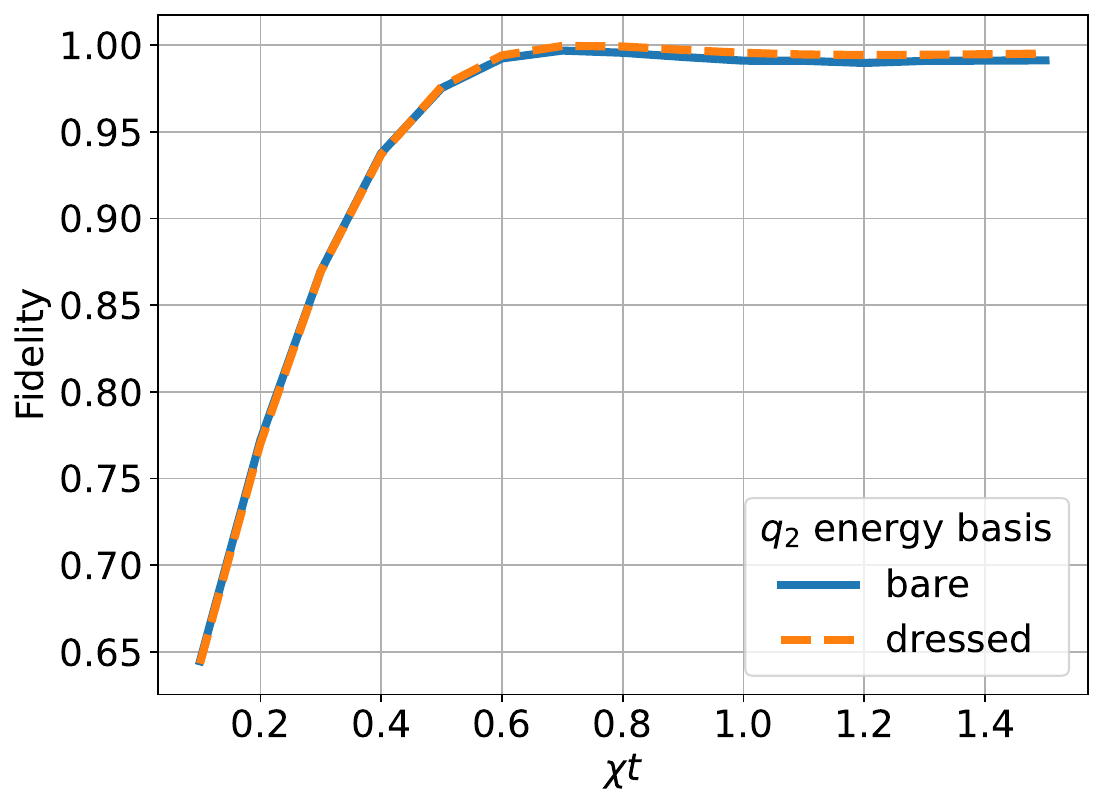}
    \put(-0.93\columnwidth, 165){\textbf{(a)}}\\
    \includegraphics[width=0.93\columnwidth]{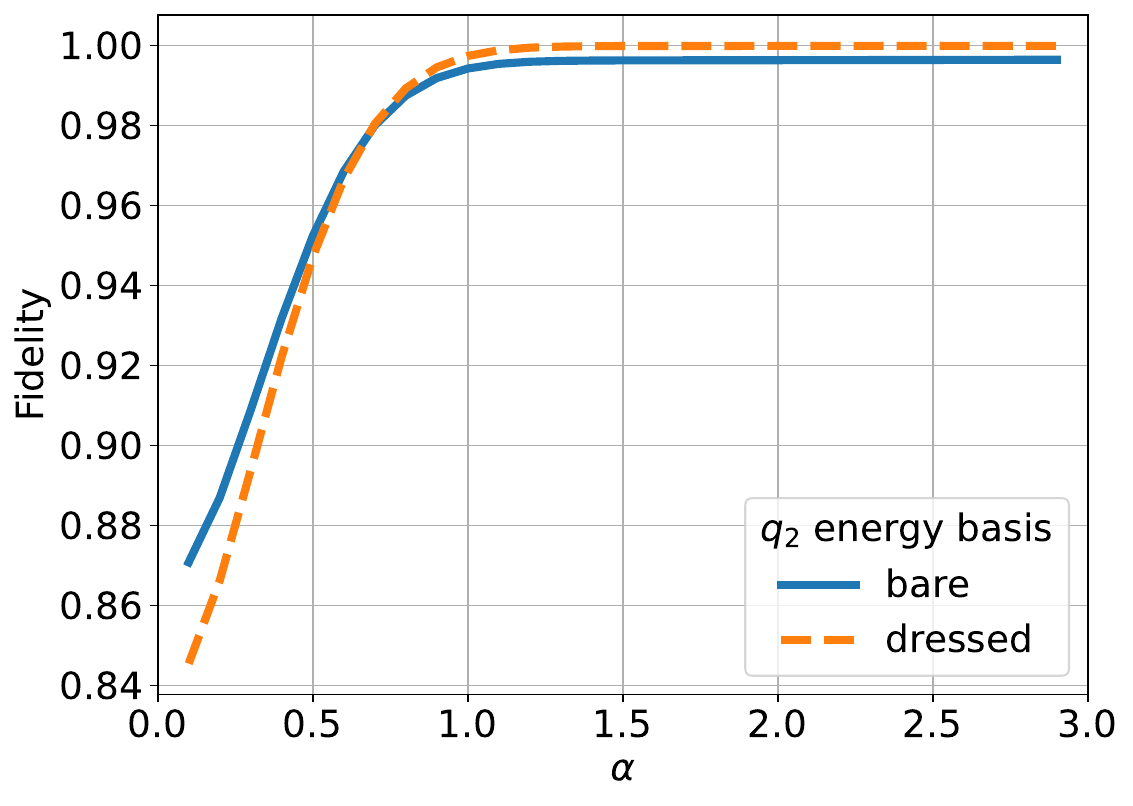}
    \put(-0.93\columnwidth, 165){\textbf{(b)}}\\
    \includegraphics[width=0.93\columnwidth]{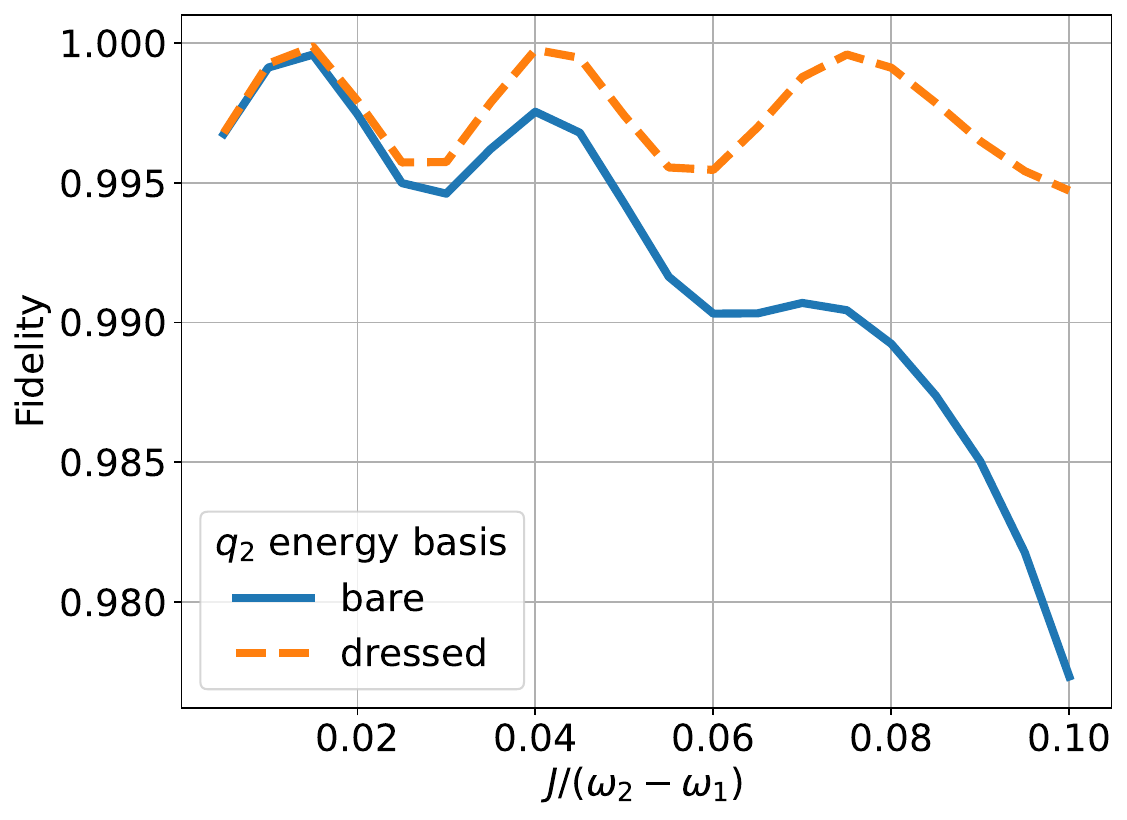}
    \put(-0.93\columnwidth, 165){\textbf{(c)}}\\
    \caption{Fidelity in bare and dressed bases calculated using the Hamiltonian expressed in the $q_2$ energy basis. Parameters are the same as in Fig.~\ref{fig:2qo_fq2q1q2}.
    }
    \label{fig:2qo_fq2bd}
\end{figure}

The fidelity in both bare and dressed bases is numerically investigated as a function of $\chi t$, $\alpha$, and $J/(\omega_{2}-\omega_{1})$, as shown in Figs.~\ref{fig:2qo_fq2bd}(a)-(c) respectively.
Figure~\ref{fig:2qo_fq2bd}(a) shows that, for very short measurement times, the fidelity is low in both bases. At a measurement time of $t \approx t_d/2$, fidelity reaches its maximum value in both bases. Beyond this point, the fidelity in the dressed basis remains higher than that in the bare basis. However, in both bases, the fidelity gradually decreases as the measurement time further increases.
As illustrated in Fig.~\ref{fig:2qo_fq2bd}(b), for small values of $\alpha$, the bare basis performs better. A basis crossover occurs around $\alpha\approx0.7$. Beyond this point, the fidelity in the dressed basis is higher. This implies that at a fixed measurement time (here, $t_d$), measuring gradually becomes better suited to the dressed basis as $\alpha$ increases.
Figure~\ref{fig:2qo_fq2bd}(c) shows how the coupling strength between two flux qubits affects the fidelity. For the chosen parameters ($t=t_d$ and $\alpha=1$), the dressed basis yields a higher fidelity than the bare basis. Additionally, the coupling strength has a larger impact on fidelity in the bare basis compared to the dressed basis.

Our numerical simulations reveal that, for small $\alpha$, the bare basis yields higher fidelity compared to the dressed basis. However, the bare basis is more sensitive to the coupling strength between the flux qubits, demonstrating a significant decline in measurement quality as the coupling strength increases. Conversely, for large $\alpha$, the dressed basis offers higher fidelity, characterized by oscillations with $J$. Notably, the average fidelity across these oscillations exhibits only a slight decrease, suggesting that these may be transient oscillations. This transient nature is possibly due to the non-adiabatic change in the operating point and the presence of multiple energy levels in the system.

\subsubsection{Energy basis of both flux qubits}

This section analyzes the Hamiltonian operator in the $q_1$-$q_2$ energy basis (Eq.~\eqref{equ:ht3_q1q2}). Due to its complexity, we adopt an analytical approach by focusing on two specific interaction regimes between two flux qubits: flux-dominated and tunneling-dominated regimes.

Within the regime defined by
\begin{equation*}
    \{\abs{\cos(\theta_1)}, \abs{\cos(\theta_{\text{eff,2}})}\} \gg  \{\abs{\sin(\theta_1)}, \abs{\sin(\theta_{\text{eff,2}})}\},
    \label{bed:c12gs}
\end{equation*}
we can approximate the interaction term between $q_1$ and $q_2$ as
$J_{zz}\tilde{\hat{\sigma}}_1^z\tilde{\hat{\sigma}}_2^z$ with $J_{zz}=J\cos(\theta_1)\cos(\theta_{\text{eff,2}})$, leading to
\begin{align}
    \begin{split}
        \tilde{\hat{H}}^{(e)}_{zz}\approx-\frac{\hat{\delta}_{\text{eff2,n}}}{2}\tilde{\hat{\sigma}}_2^z+J_{zz} \tilde{\hat{\sigma}}_1^z\tilde{\hat{\sigma}}_2^z.
    \end{split}
    \label{equ:Ht2_zz_q1q2}
\end{align}
We refer to this approximation as the $zz$ approximation. Under this approximation, $\tilde{\hat{H}}^{(3)}_{zz}$ becomes a diagonal matrix, implying that the $q_1$-$q_2$ energy basis already coincides with the dressed basis. Consequently, the criterion for a valid RWA is automatically fulfilled. This signifies that the $zz$ approximation tends to yield higher fidelity in the regime studied.

\begin{figure}[tb]
    \centering
    \includegraphics[width=0.93\columnwidth]{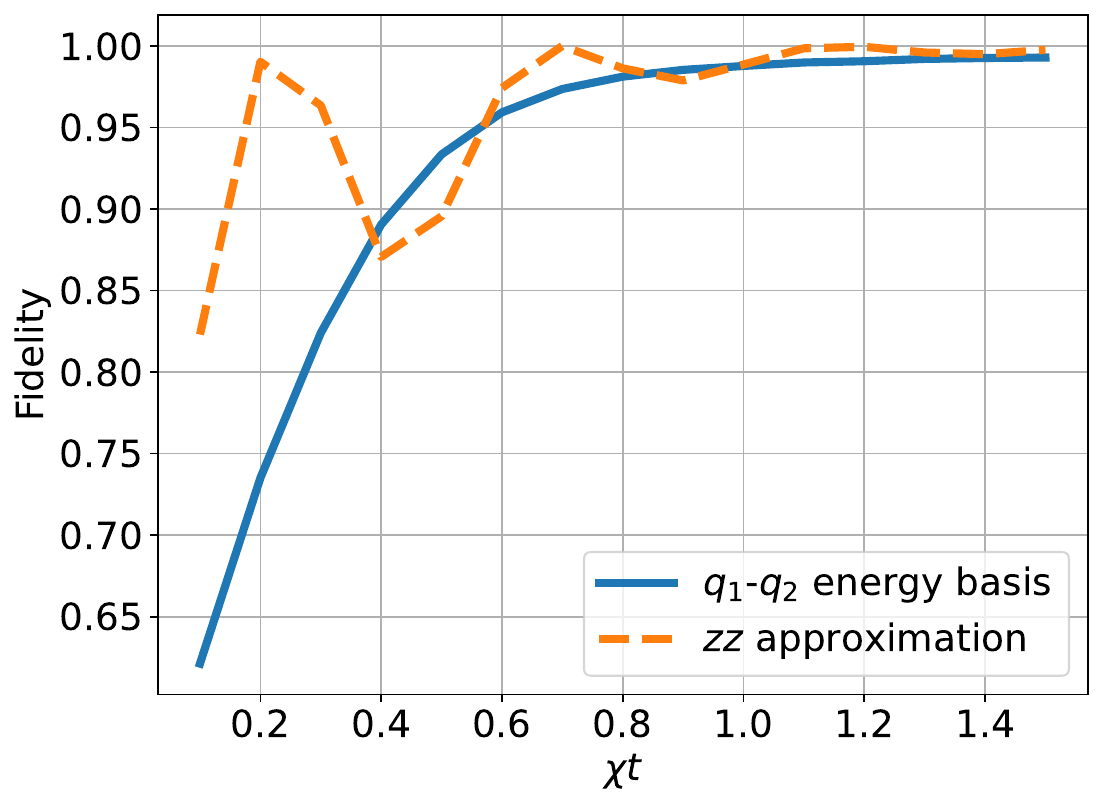}
    \put(-0.93\columnwidth, 165){\textbf{(a)}}\\
    \includegraphics[width=0.93\columnwidth]{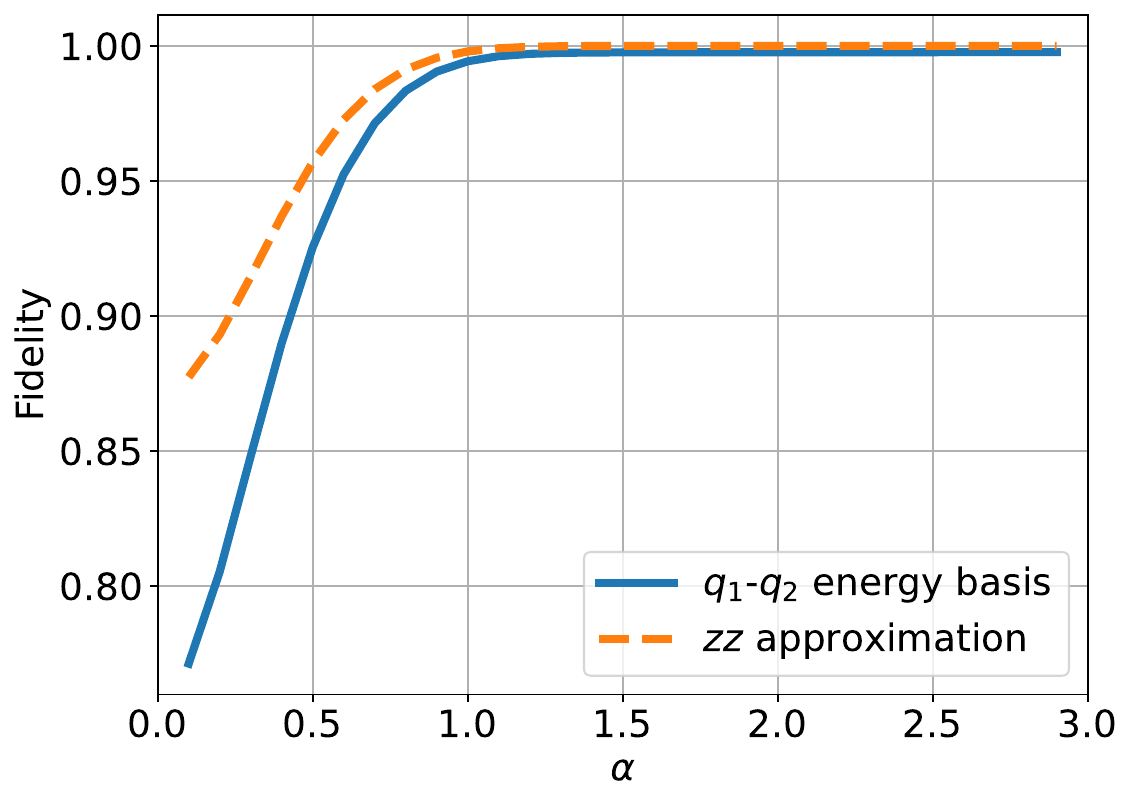}
    \put(-0.93\columnwidth, 165){\textbf{(b)}}\\
    \includegraphics[width=0.93\columnwidth]{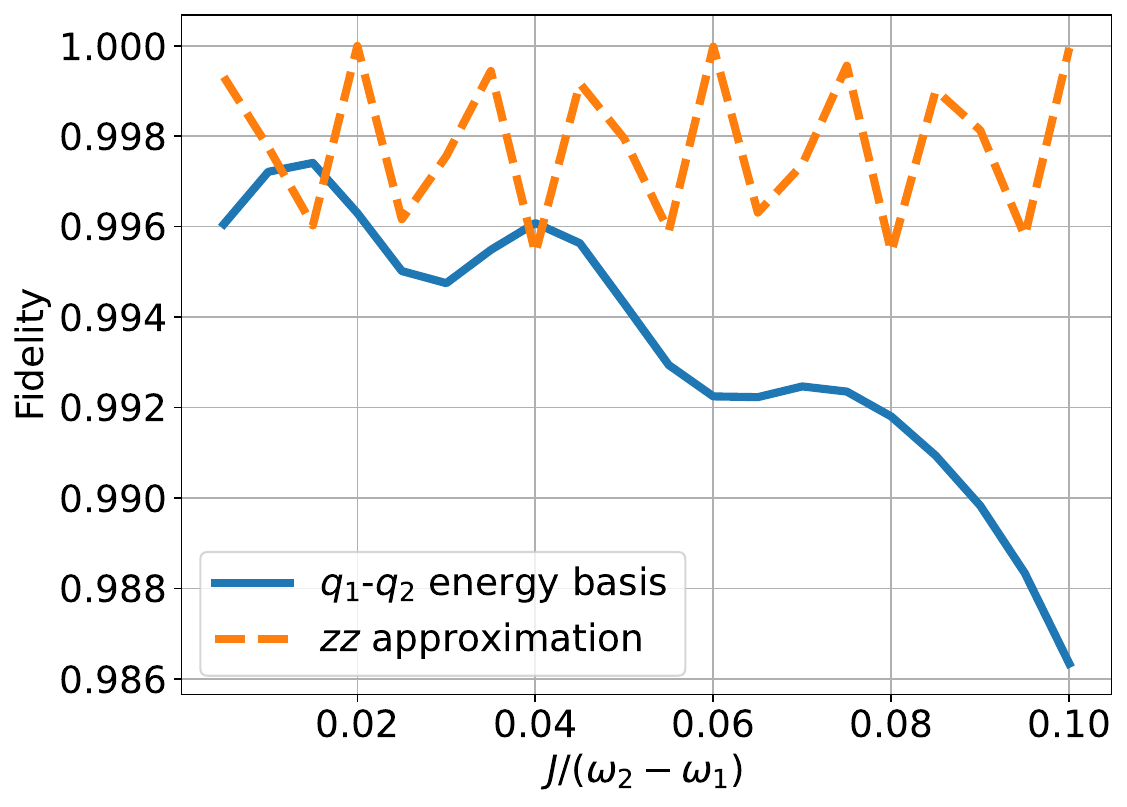}
    \put(-0.93\columnwidth, 165){\textbf{(c)}}\\
    \caption{Fidelity calculated using the Hamiltonian expressed in the $q_1$-$q_2$ energy basis without approximation and under $zz$ approximation. Parameters are the same as in Fig.~\ref{fig:2qo_fq2q1q2}.
    }
    \label{fig:2qo_fq1q2zz}
\end{figure}

Figure~\ref{fig:2qo_fq1q2zz} compares the state fidelity obtained from the $q_1$-$q_2$ energy-basis Hamiltonian with that from the flux-dominant $zz$ approximation. The key distinction is whether the effective qubit-qubit coupling remains QND throughout the measurement. In the full Hamiltonian, the rotated interaction includes transverse and noncommuting components that drive coherent entanglement between target qubit $q_2$ and spectator qubit $q_1$, inducing non-QND dynamics. Consequently, fidelity increases monotonically with $\chi t$ and saturates only when the measurement strength, depending on $\chi t$ and $\alpha$, dominates this neighbor-induced dynamics. Under $zz$ approximation, the coupling reduces to the purely longitudinal term $J_{zz}\tilde{\sigma}_1^z\tilde{\sigma}_2^z$, which is QND-compliant. The spectator qubit then induces only conditional phase shifts, enabling high fidelity already at shorter $\chi t$ and producing damped oscillations due to the photon number resolved POVM, as shown in Fig.~\ref{fig:2qo_fq1q2zz}(a).
The advantage of the $zz$ approximation is most pronounced at small $\alpha$, where the measurement is most sensitive to transverse mixing, and vanishes for $\alpha\gtrsim 1$ in the near-projective regime (Fig.~\ref{fig:2qo_fq1q2zz}(b)). Increasing $J/(\omega_2-\omega_1)$ degrades fidelity only in the $q_1$-$q_2$ energy-basis, as it amplifies non-QND transverse terms and hence neighbor-induced basis rotation, whereas $zz$ approximation remains robust near $0.998$, exhibiting only mild phase-induced oscillations (Fig.~\ref{fig:2qo_fq1q2zz}(c)). These results suggest a clear design guideline for coupled flux-qubit readout: bias the qubits into the flux-dominant regime during measurement, or choose sufficiently strong $\chi t$ and $\alpha$ such that dispersive interaction dominates, thereby suppressing neighbor-induced errors.

We now explore the tunneling-dominated regime, characterized by
\begin{equation*}
\{\abs{\cos(\theta_1)}, \abs{\cos(\theta_{\text{eff,2}})}\} \ll \{\abs{\sin(\theta_1)}, \abs{\sin(\theta_{\text{eff,2}})}\}.
\end{equation*}
In this regime, the interaction between flux qubits simplifies to the $xx$ approximation
$J_{xx}\tilde{\hat{\sigma}}_1^x\tilde{\hat{\sigma}}_2^x$ with $J_{xx}=J\sin(\theta_1)\sin(\theta_{\text{eff,2}})$, resulting in the Hamiltonian
\begin{equation}
    \tilde{\hat{H}}^{(e)}_{xx}\approx-\frac{\hat{\delta}_{\text{eff2,n}}}{2}\tilde{\hat{\sigma}}_2^z+J_{xx} \tilde{\hat{\sigma}}_1^x\tilde{\hat{\sigma}}_2^x.
\label{equ:Hq4_xx}
\end{equation}
Diagonalizing $\tilde{H}^{(e)}_{xx}$ defines the dressed basis (see Appendix~\ref{app:seq_q1_q2_xx}).
The valid RWA condition for dressed basis is then given by
\begin{equation*}
    \left\Vert \frac{J_{xx}~\chi \left(\hat{a}^{\dag}\hat{a}+\frac{1}{2}\right)}{\hat{\delta}_{\text{eff2,n}} \delta_{\text{eff,2}}/4 + J_{xx}^2}\right\Vert  \ll 1.
\end{equation*}
In comparison, the valid RWA condition for the bare basis is
\begin{equation*}
    \left\Vert \frac{J_{xx}}{\hat{\delta}_{\text{eff2,n}}/2} \right\Vert \ll 1.
\end{equation*}
If the condition
\begin{equation*}
    \left\Vert \chi \left(\hat{a}^{\dag}\hat{a}+\frac{1}{2}\right) \right\Vert=\sqrt{\left(\delta_{\text{eff,2}}/2\right)^2+J_{xx}^2}
    \label{bed:q1q2xx}
\end{equation*}
is satisfied, there is no distinction between bare and dressed bases.
Consequently, this provides the criterion for a basis crossover in the $q_1$-$q_2$ energy basis representation in the tunneling-dominated regime.

\begin{figure}[tb]
    \centering
    \includegraphics[width=0.91\columnwidth]{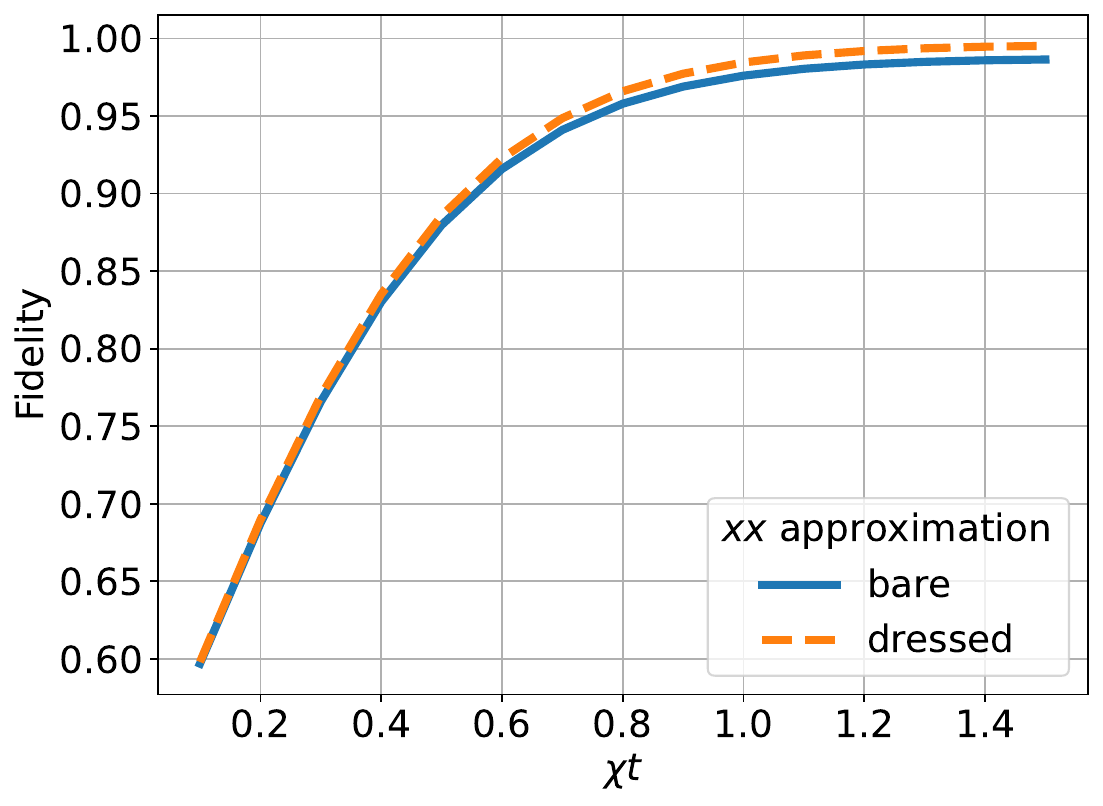}
    \put(-0.93\columnwidth, 160){\textbf{(a)}}\\
    \includegraphics[width=0.92\columnwidth]{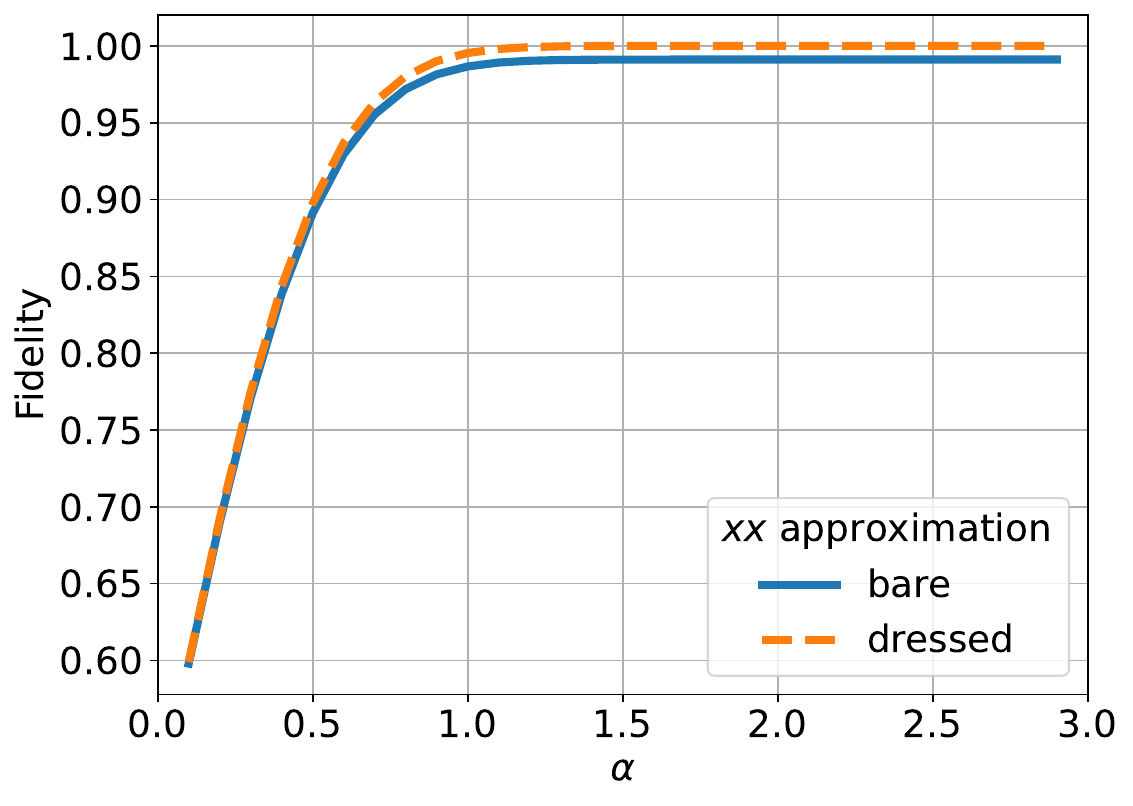}
    \put(-0.93\columnwidth, 158){\textbf{(b)}}\\
    \includegraphics[width=0.91\columnwidth]{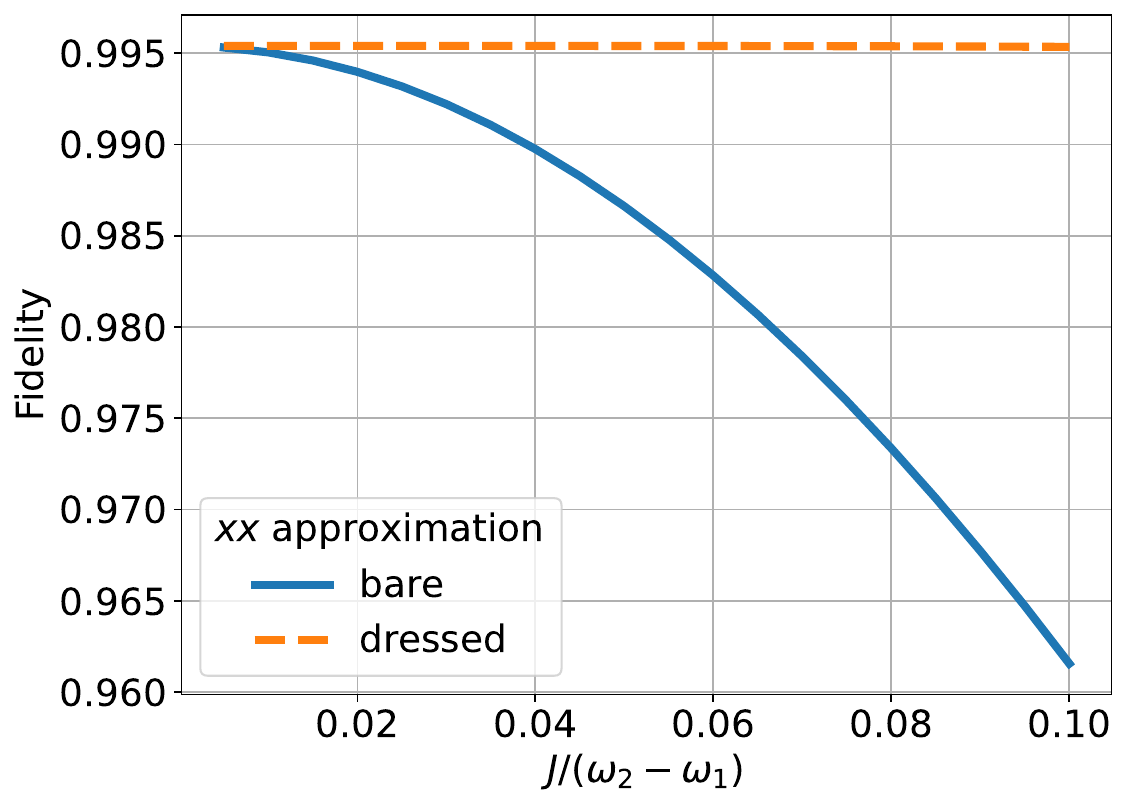}
    \put(-0.93\columnwidth, 156){\textbf{(c)}}\\
    \vspace{-.5em}
    \caption{Fidelity in bare and dressed bases calculated using the Hamiltonian expressed in the $q_1$-$q_2$ energy basis under the $xx$ approximation with $N=27$, $\eta=1.25$, $\Delta_{2}/\epsilon_2 = 10$, and $\delta/g=8$. Parameters are: (a) $J/(\omega_{2}-\omega_{1})=0.05$ and $\alpha=1$; (b) $J/(\omega_{2}-\omega_{1})=0.05$ and $t = t_d$; and (c) $t = t_d$ and $\alpha=1$.
    }
    \label{fig:2qo_q1q2xx}
\end{figure}

Figure~\ref{fig:2qo_q1q2xx} presents results of a numerical investigation of fidelity using a Hamiltonian under $xx$ approximation, comparing bare and dressed bases.
As shown in Fig.~\ref{fig:2qo_q1q2xx}(a), maximum fidelity is achieved in both bases around a measurement time of $t \approx t_d$. However, for $t>t_d/\pi$, the dressed basis surpasses the bare basis. A similar trend is observed in Fig.~\ref{fig:2qo_q1q2xx}(b). When the amplitude of the resonator coherent state is relatively small, $\alpha\lesssim0.5$, the fidelity in both bases shows minimal difference. As $\alpha$ increases, the dressed basis exhibits both higher fidelity and saturation value. Figure~\ref{fig:2qo_q1q2xx}(c) demonstrates that increasing the coupling strength negatively impacts the fidelity in the bare basis. Conversely, fidelity in a dressed basis remains consistently high, demonstrating robustness against changes in coupling strength.
The results highlight the advantage of a dressed basis over a bare basis. Specifically, while the bare basis exhibits high sensitivity to changes in coupling strength, the dressed basis exhibits remarkable resilience to such perturbations and maintains a consistently high fidelity.

To summarize, we have investigated the condition for a valid RWA within the $q_1$-$q_2$ energy basis by analyzing two scenarios, including $zz$ and $xx$ approximations. Under the $zz$ approximation, the Hamiltonian already exists in a diagonalized form, automatically satisfying the RWA condition. This simplifies analysis and guarantees higher fidelity. In contrast, the $xx$ approximation necessitates a specific condition for a valid RWA. Employing a dressed basis offers potential advantages in terms of fidelity and robustness.

\subsection{Simultaneous measurement}

\begin{figure}[t]
    \centering
    \includegraphics[width=\columnwidth]{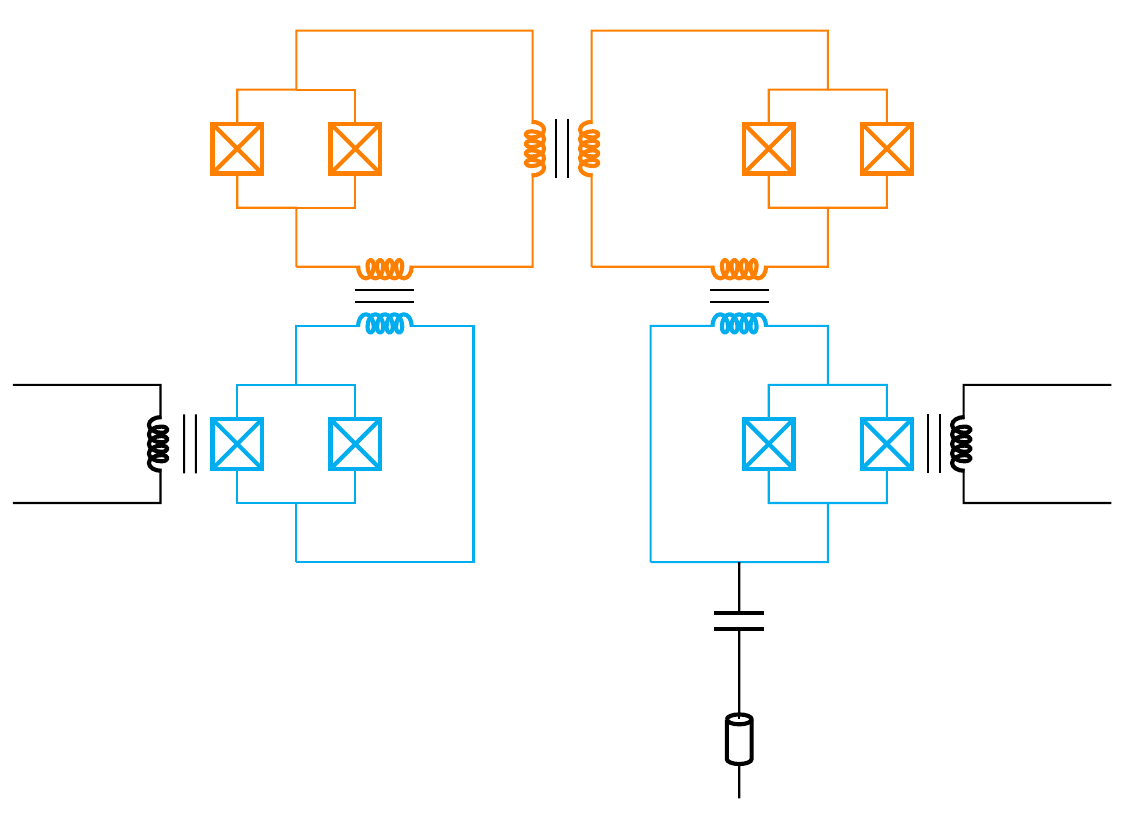}
    \put(-182,176){Flux qubit $q_1$}
    \put(-120,176){Flux qubit $q_2$}
    \put(-168,45){QFP\textsubscript{1}}
    \put(-78,45){QFP\textsubscript{2}}
    \put(-80,18){Resonator}
    \caption{Simultaneous measurement model. Each flux qubit signal is coupled with a QFP. Through adiabatic annealing, the flux qubit state is latched into the corresponding QFP. Then, QFP\textsubscript{2} is read out using a resonator.}
    \label{fig:modell2m}
\end{figure}
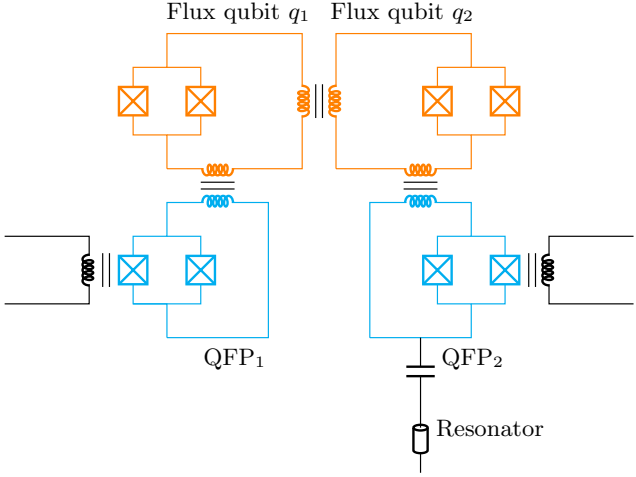

This section explores the simultaneous measurement model, in which both QFP\textsubscript{1} and QFP\textsubscript{2} undergo adiabatic annealing, as shown in Fig.~\ref{fig:modell2m}. Then, the state of QFP\textsubscript{2} is measured using a resonator. In contrast to the sequential model, where one qubit is actively measured while the other is assumed to be static, the simultaneous measurement model keeps both qubits dynamically involved. After annealing, the states of flux qubits $q_1$ and $q_2$ are latched into their respective QFP\textsubscript{1} and QFP\textsubscript{2}. The primary focus is on reading out the state of QFP\textsubscript{2} via its associated resonator, while any interaction of QFP\textsubscript{1} with a potentially connected resonator is neglected. This simplification is valid because once latched, QFP\textsubscript{1}’s state no longer influences QFP\textsubscript{2} or the resonator connected to QFP\textsubscript{2}. To ensure this isolation, we assume negligible interactions between QFPs, achieved through an optimized design featuring sufficient spatial separation, minimized mutual inductance, and distinct operational frequencies to effectively suppress crosstalk. Consequently, the system is characterized by a Hamiltonian expressed in the flux basis
\begin{eqnarray}
    \hat{H}&=&-\frac{1}{2}\left(\epsilon_{1} \hat{\sigma}_1^z+\Delta_{\text{eff,1}}\hat{\sigma}_1^x\right)-\frac{1}{2}\left(\epsilon_{2} \hat{\sigma}_2^z+\Delta_{\text{eff,2}}\hat{\sigma}_2^x\right)\nonumber\\
    & &+J\hat{\sigma}_1^z\hat{\sigma}_2^z
    +g_2\hat{\sigma}_2^z\left(\hat{a}^{\dag}+\hat{a}\right)+\omega_r \hat{a}^{\dag}\hat{a},
    \label{eq:hamil_simul_measu}
\end{eqnarray}
where $\epsilon_{1}$ and $\epsilon_{2}$ are energies of $q_1$ and $q_2$, respectively, $\Delta_{\text{eff,1}}$ and $\Delta_{\text{eff,2}}$ are the corresponding effective detunings, $\omega_r$ is resonator frequency, $J$ denotes coupling strength between two flux qubits, and $g_2$ corresponds to coupling strength between QFP\textsubscript{2} and resonator.

Transforming Eq.~\eqref{eq:hamil_simul_measu} to the $q_1$-$q_2$ energy basis and applying the standard displacement and dispersive-elimination steps yields the effective Hamiltonian (more details see Appendix~\ref{app:sim_q1_q2_energy})
\begin{eqnarray}
    \tilde{\hat{H}}_{(e)}&=&-\frac{\delta_{\text{eff,1}}}{2}\tilde{\hat{\sigma}}_1^z-\frac{\hat{\delta}_{\text{eff2,n}}}{2}\tilde{\hat{\sigma}}_2^z +J\left[\cos(\theta_{\text{eff,1}}) \tilde{\hat{\sigma}}_1^z - \right.\nonumber\\
    & &\left.\sin(\theta_{\text{eff,1}})\tilde{\hat{\sigma}}_1^x\right]
    \left[\cos(\theta_{\text{eff,2}}) \tilde{\hat{\sigma}}_2^z-\sin(\theta_{\text{eff,2}}) \tilde{\hat{\sigma}}_2^x\right],\qquad
    \label{equ:ht3_q1q1_b}
\end{eqnarray}
where $\delta_{\text{eff,i}}=\omega_{\text{eff,i}}-\omega_r$ and $\theta_{\text{eff,i}}$ is defined as $\tan(\theta_{\text{eff,i}})=\Delta_{\text{eff,i}}/\epsilon_{i}$ with $\omega_{\mathrm{eff},i}=\sqrt{\epsilon_i^2+\Delta_{\mathrm{eff},i}^2}$ for $i\in\{1,2\}$, and $\hat{\delta}_{\text{eff2,n}}=\delta_{\text{eff,2}}+\chi\left(2 \hat{n} +1\right)$ with $\chi=g_{2}^2\sin^2(\theta_{\text{eff,2}})/\delta_{\text{eff,2}}$.
Transforming back into the flux basis yields
\begin{eqnarray}
    \hat{H}_{(f)}&=&-\frac{\delta_{\text{eff,1}}}{2}\left[\cos(\theta_{\text{eff,1}}) \hat{\sigma}_1^z+\sin(\theta_{\text{eff,1}}) \hat{\sigma}_1^x\right]+J\hat{\sigma}_1^z\hat{\sigma}_2^z\nonumber\\
    & &-\frac{\hat{\delta}_{\text{eff2,n}}}{2}\left[\cos(\theta_{\text{eff,2}}) \hat{\sigma}_2^z+\sin(\theta_{\text{eff,2}}) \hat{\sigma}_2^x\right].
    \label{equ:h2_q1q2_fb}
\end{eqnarray}
Similar to the previous analysis, we consider flux-dominated and tunneling-dominated regimes.

\subsubsection{Flux-dominated regime}

We first analyze the Hamiltonian within the flux-dominated regime, assuming
\begin{equation*}
    \{\abs{\cos(\theta_{\text{eff,1}})}, \abs{\cos(\theta_{\text{eff,2}})}\} \gg \{\abs{\sin(\theta_{\text{eff,1}})}, \abs{\sin(\theta_{\text{eff,2}})}\}.
\end{equation*}
Under this condition, the interaction between $q_1$ and $q_2$ can be approximated as $J_{zz}\tilde{\hat{\sigma}}_1^z\tilde{\hat{\sigma}}_2^z$ with $J_{zz}=J\cos(\theta_{\text{eff,1}})\cos(\theta_{\text{eff,2}})$, which we refer to as $zz$ approximation. This approximation leads to a simplified Hamiltonian
\begin{equation}
    \tilde{\hat{H}}^{(3)}_{zz}\approx-\frac{\delta_{\text{eff,1}}}{2}\tilde{\hat{\sigma}}_1^z-\frac{\hat{\delta}_{\text{eff2,n}}}{2}\tilde{\hat{\sigma}}_2^z
    +J_{zz} \tilde{\hat{\sigma}}_1^z\tilde{\hat{\sigma}}_2^z.
    \label{ht33_zz}
\end{equation}

\begin{figure}[tb]
    \centering
    \includegraphics[width=0.92\columnwidth]{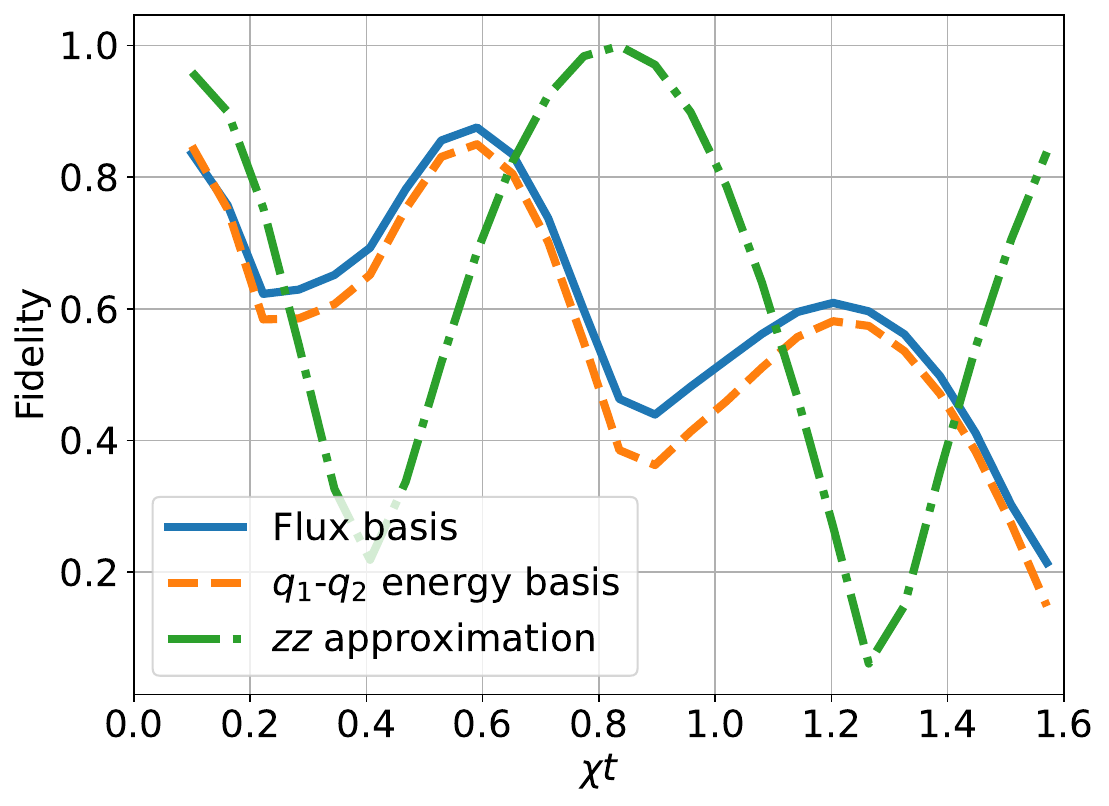}
    \put(-0.93\columnwidth, 160){\textbf{(a)}}\\
    \includegraphics[width=0.92\columnwidth]{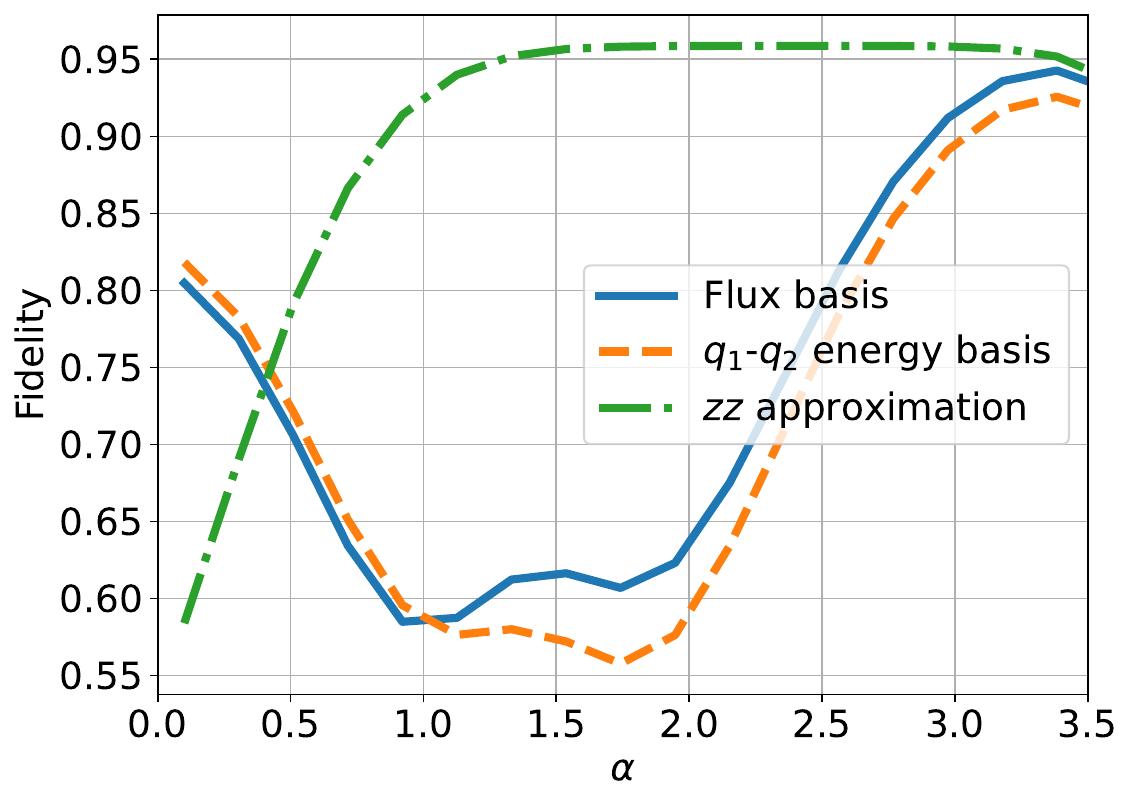}
    \put(-0.93\columnwidth, 156){\textbf{(b)}}\\
    \includegraphics[width=0.92\columnwidth]{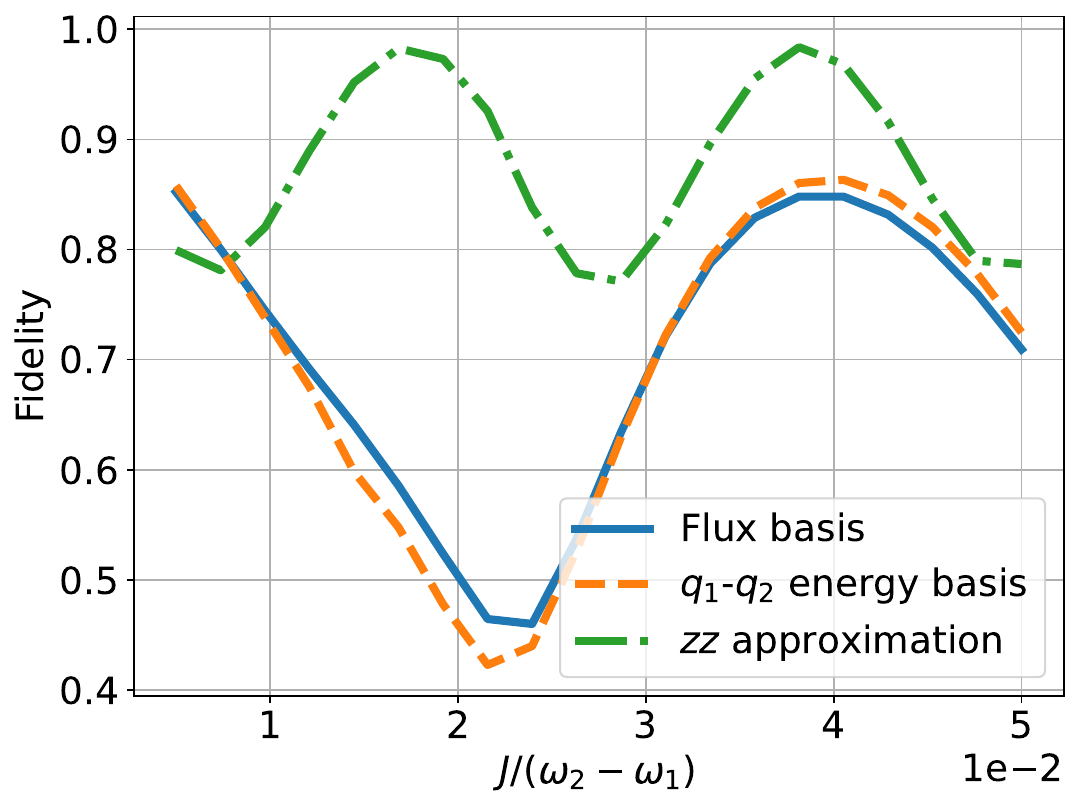}
    \put(-0.93\columnwidth, 162){\textbf{(c)}}\\
    \vspace{-.5em}
    \caption{Fidelity calculated using the Hamiltonian expressed in the flux basis, the $q_1$-$q_2$ energy basis without approximation, and under $zz$ approximation with $N=21$, $\eta=1.25$, $\Delta_{2}/\epsilon_2 = 0.5$, and $\delta/g=8$. Parameters are: (a) $J/(\omega_{2}-\omega_{1})=0.05$ and $\alpha=2$; (b) $J/(\omega_{2}-\omega_{1})=0.05$ and $t=t_d/2$; and (c) $t=t_d/2$ and $\alpha=1$.
    }
    \label{fig:2qm_q1q2zz}
\end{figure}

Figure~\ref{fig:2qm_q1q2zz} compares the fidelity of measuring $q_2$ in the simultaneous-measurement model under three different effective Hamiltonians: the flux basis, the $q_1$-$q_2$ energy basis, and the flux-dominant $zz$ approximation. The photon-number dependent shift $\hat{\delta}_{\mathrm{eff},2,n}$ together with spectator-conditioned detuning imprint basis- and state-dependent phases that are mapped into oscillatory behavior by the half-plane POVM.
Under $zz$ approximation, $J$ acts primarily as a conditional detuning, producing short-$\chi t$ windows of high fidelity (Fig.~\ref{fig:2qm_q1q2zz}(a)) and phase-induced oscillations with $\chi t$ and $J$ (Figs.~\ref{fig:2qm_q1q2zz}(a) and (c)), whereas in the flux and full energy-basis transverse noncommuting components can rotate the effective measurement basis and induce mixing during measurement, smearing the extrema and degrading fidelity with increasing $J$. Moreover, changing $\alpha$ reweights photon-number sectors and hence the phase-averaging that enters the POVM mapping, shifting the location of the optimal operating point (Fig.~\ref{fig:2qm_q1q2zz}(b)). These results show that simultaneous measurement accentuates the competition between measurement-induced localization and coherent hybridization, implying an operational optimum in $(t,\alpha)$ that depends on $J$ and favoring longitudinal operation during readout or parameters where measurement dominates hybridization for robust single-qubit discrimination of $q_2$.

\subsubsection{Tunneling-dominated regime}

This section explores the tunneling-dominated regime under the condition
\begin{equation*}
    \{\abs{\cos(\theta_{\text{eff,1}})}, \abs{\cos(\theta_{\text{eff,2}})}\} \ll \{\abs{\sin(\theta_{\text{eff,1}})}, \abs{\sin(\theta_{\text{eff,2}})}\}.
\end{equation*}
Within this regime, we approximate the interaction term between $q_1$ and $q_2$ as $J_{xx}\tilde{\hat{\sigma}}_1^x\tilde{\hat{\sigma}}_2^x$  (the $xx$ approximation), where $J_{xx}=J \sin(\theta_\text{eff,1})\sin(\theta_{\text{eff,2}})$, leading to the simplified Hamiltonian
\begin{equation}
    \tilde{\hat{H}}^{(3)}_{xx}=-\frac{\delta_{\text{eff,1}}}{2}\tilde{\hat{\sigma}}_1^z-\frac{\hat{\delta}_{\text{eff2,n}}}{2}\tilde{\hat{\sigma}}_2^z+J_{xx} \tilde{\hat{\sigma}}_1^x \tilde{\hat{\sigma}}_2^x.
\label{equ:ht3_xx_b}
\end{equation}

Analogously, we obtain the condition for good RWA in the dressed basis through the diagonalization of the Hamiltonian
\begin{equation*}
    \left\Vert \frac{J_{xx}\left[\left(\delta_{\text{eff,2}}\pm \delta_{\text{eff,1}}\right)\pm\left(\hat{\delta}_{\text{eff2,n}}\pm \delta_{\text{eff,1}}\right)\right]/2}{\left(\delta_{\text{eff,2}}\pm \delta_{\text{eff,1}}\right)\left(\hat{\delta}_{\text{eff2,n}} \pm\delta_{\text{eff,1}}\right)/4 +J_{xx}^2} \right\Vert \ll 1,
\end{equation*}
where $\theta_{n\pm }$ is defined as $\tan(\theta_{n\pm })=2J_{xx}/\left(\delta_{\text{eff2,n}}\pm \delta_{\text{eff,1}}\right)$. In the bare basis, the condition becomes
\begin{equation*}
    \left\Vert \frac{J_{xx}}{\left(\hat{\delta}_{\text{eff2,n}}\pm \delta_{\text{eff,1}}\right)/2} \right\Vert \ll 1.
\end{equation*}

The equation for the crossover point between bare and dressed bases is given by
\begin{equation*}
    \left\Vert\chi\left(\hat{a}^\dagger\hat{a}+\frac{1}{2}\right)\right\Vert = \sqrt{\left(\delta_{\text{eff,2}}- \delta_{\text{eff,1}}\right)^2/4 + J_{xx}^2}.
    \label{bed:2qmxx2}
\end{equation*}

\begin{figure}[tb]
    \centering
    \includegraphics[width=0.92\columnwidth]{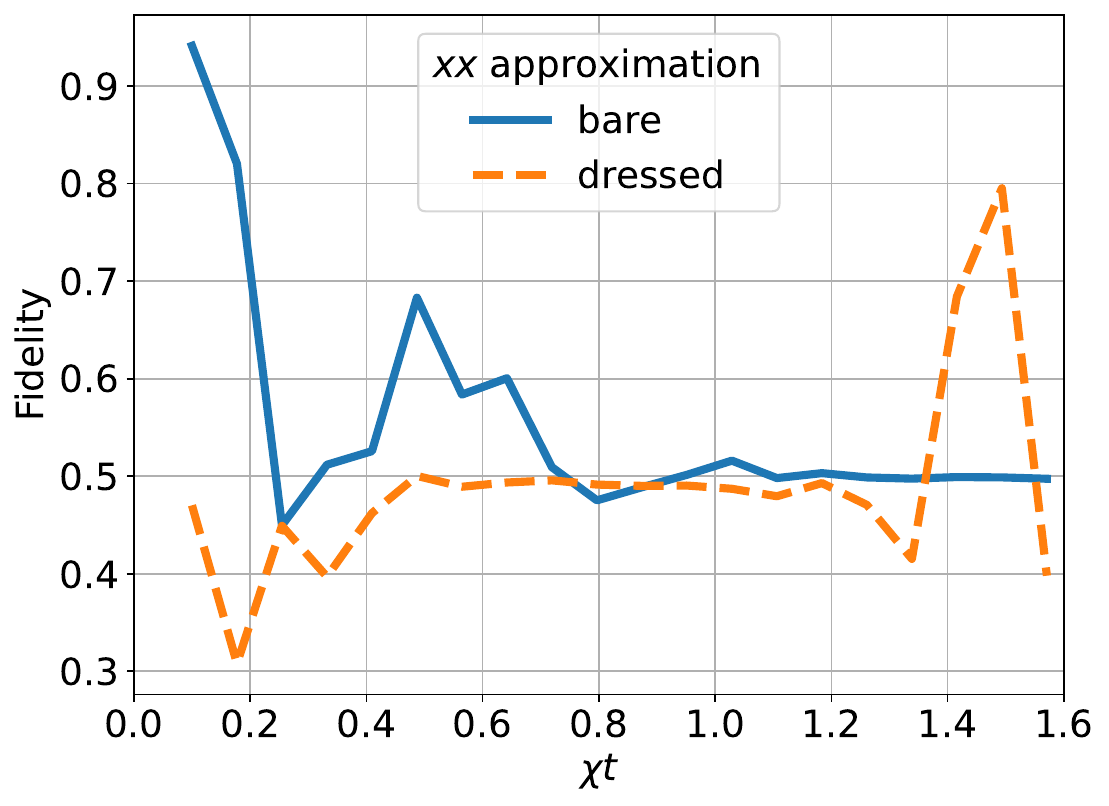}
    \put(-0.93\columnwidth, 160){\textbf{(a)}}\\
    \vspace{-2pt}
    \includegraphics[width=0.90\columnwidth]{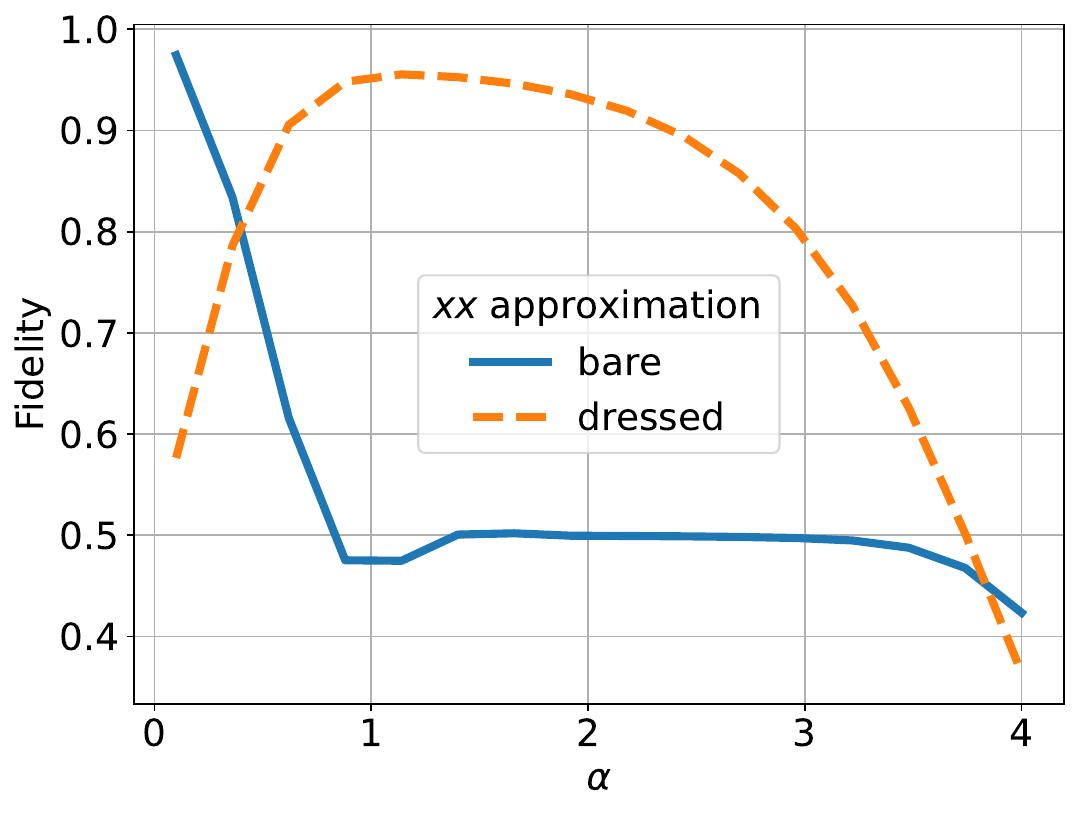}
    \put(-0.93\columnwidth, 160){\textbf{(b)}}\\
    \vspace{-2pt}
    \includegraphics[width=0.90\columnwidth]{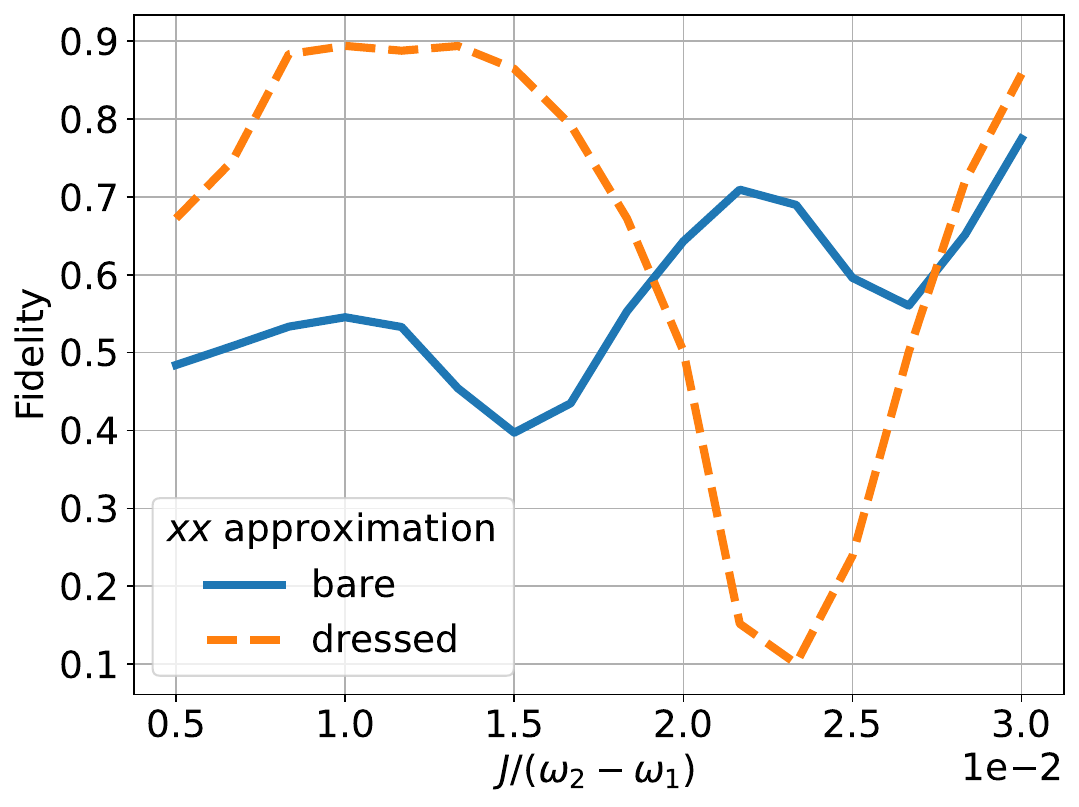}
    \put(-0.93\columnwidth, 160){\textbf{(c)}}\\
    \vspace{-.5em}
    \caption{Fidelity in bare and dressed bases calculated using the Hamiltonian expressed in the $q_1$-$q_2$ energy basis under the $xx$ approximation with $N=21$, $\eta=1$, $\Delta_{2}/\epsilon_2 = 8$, and $\delta/g=8$. Parameters are: (a) $J/(\omega_{2}-\omega_{1})=0.05$ and $\alpha=2$; (b) $J/(\omega_{2}-\omega_{1})=0.05$ and $t = t_d$; and (c) $t = t_d$ and $\alpha=0.5$.
    }
    \label{fig:2qm_q1q2xx}
\end{figure}

Figure~\ref{fig:2qm_q1q2xx} presents numerical fidelities in the tunneling-dominated regime, where the effective interaction is transverse and therefore does not commute with the measured observable on $q_2$, enabling coherent hybridization during discrimination. This non-QND dynamics produces the basis crossover in Fig.~\ref{fig:2qm_q1q2xx}(a). At short measurement times, the readout detects $q_2$ before significant hybridization develops, yielding higher fidelity in the bare basis, whereas at longer times the coupled system evolves toward joint eigenmodes and the dressed basis becomes more stable. The $\alpha$ dependence in Fig.~\ref{fig:2qm_q1q2xx}(b) follows from the same mechanism. Varying $\alpha$ reweights photon-number sectors and hence the effective detuning $\hat{\delta}_{\mathrm{eff},2,n}$ that controls hybridization, yielding nonmonotonic behavior and shifting the balance between bare- and dressed-basis performance. Any additional decrease at the largest $\alpha$ is consistent with increased sector averaging or approaching dispersive limits. Figure~\ref{fig:2qm_q1q2xx}(c) shows that varying $J/(\omega_2-\omega_1)$ changes the hybridization rate and dressed-mode splitting, leading to oscillatory fidelity and a clear crossover, with the dressed basis responding more strongly because it tracks the hybridized eigenmodes. The results demonstrate that in the tunneling-dominated regime the optimal measurement basis is an operational choice set by $(t,\alpha,J)$: fast discrimination can realize high-fidelity bare-basis readout, whereas longer integration favors dressed-basis readout once hybridization becomes appreciable.

\section{Discussion}\label{sec:discu}

This section discusses practical integration constraints, the impact of relaxation and dephasing, and the expected noise robustness of the proposed bases and protocols.

\subsection{Experimental considerations}

In an experimental implementation, three practical factors are central. First, because the QFP is a flux-controlled rf-SQUID that mediates an effective mutual inductance between the flux qubit and the readout resonator, QFP-QFP crosstalk can arise from static loop-loop mutual inductance and cross-coupling between bias lines in the flux-bias network \cite{grover2020fast}.
In practice, these parasitic couplings are characterized as a flux-crosstalk matrix and compensated by applying the inverse calibration map. Model-independent iterative calibration has been demonstrated on devices with up to 27 control loops with crosstalk-coefficient errors at the sub-percent level \cite{dai2021calibration}, and multi-$Z$-line cancellation-matrix approaches can suppress residual crosstalk to the $\sim 10^{-4}$ level in qubit-coupler systems \cite{aguila2025characterizing}.
Second, isolation during readout is dynamic. QFP-mediated readout can preserve qubit lifetime when the QFP is biased ``off", suppressing Purcell-limited decay into the readout resonator, while enabling fast, high-contrast discrimination when biased ``on" \cite{grover2020fast}. However, during the annealing (latching) ramp isolation can be limited by nonadiabatic population transfer induced by overly sharp ramps and flux-control distortions and long-timescale transients in realistic $Z$-lines, which can shift the instantaneous operating point and introduce backaction \cite{hellings2025calibrating,Aggarwal2025Mar,rol2020time}. Smooth ramps combined with in-situ transfer-function characterization and predistortion are therefore important in practice.
Finally, nonideal QFP parameters, such as junction asymmetry and $I_c$ spread, shift the balance point and modify the bifurcation threshold, reducing latching margin and requiring per-device calibration of operating biases and ramp endpoints. Related flux-qubit experiments show that junction asymmetry can induce nonlinear control crosstalk during annealing and can be mitigated by in-situ path correction \cite{Khezri2021Anneal}. At scale, these calibration requirements need to be balanced against wiring and power constraints. Scalable addressing of many flux-bias degrees of freedom and cryogenic control-plane integration have been demonstrated in superconducting technologies, including programmable flux-bias delivery schemes \cite{Johnson2010Apr} and ultra-low-power adiabatic superconductor-logic approaches for flux control and multiplexed qubit control \cite{takeuchi2023fluxcontrollers,Takeuchi2024Microwave}.

\subsection{Decoherence and noise robustness}

In a realistic device, relaxation and dephasing introduce additional errors with effects dependent on two distinct time windows: the latching window $t_{\mathrm{lat}}$, during which the QFP anneal establishes a qubit-state-dependent latching outcome, and the integration window $t_{\mathrm{int}}$, during which the latched QFP state is discriminated via the resonator.
Energy relaxation reduces assignment fidelity primarily if it occurs before the latch outcome is fixed, i.e., within $t_{\mathrm{lat}}$.
Once a successful latch has been established, subsequent qubit relaxation influences the inferred outcome only through residual qubit--QFP coupling that may destabilize the metastable QFP state. In the target operating regime the QFP retention time $\tau_{\mathrm{ret}}$, defined as the switching time of the latched state at the operating point, satisfies $\tau_{\mathrm{ret}} \gg t_{\mathrm{int}}$, so that late qubit flips do not appreciably modify the recorded latch outcome. The remaining error budget during $t_{\mathrm{int}}$ is therefore set by latch stability, resonator integration noise, and readout chain discrimination.

Dephasing degrades readout by reducing the reproducibility of the qubit-to-pointer mapping during $t_{\mathrm{lat}}$. In flux-sensitive circuits, low-frequency $1/f^{\alpha}$ flux noise acts as quasi-static bias disorder on a single shot timescale, broadening the effective latching threshold and introducing shot-to-shot variation \cite{schondorf2020flux,quintana2017observation,rower2023evolution}.
Time-dependent ramps can further convert stochastic flux noise into effective dephasing and nonadiabatic errors via the ramp filter function \cite{setiawan2021analytic,trappen2025decoherence}, while deterministic waveform distortions primarily produce systematic trajectory errors unless they drift shot-to-shot, motivating filtered and predistorted flux control \cite{hellings2025calibrating,Aggarwal2025Mar}.
If isolation from the resonator is imperfect during $t_{\mathrm{int}}$, measurement photons can induce extra dephasing and, in the presence of dephasing noise near the qubit-resonator detuning frequency, photon-enabled transitions between dressed states \cite{gambetta2006qubit,boissonneault2009dispersive,slichter2012measurement}. Residual photons present at the start of the ramp can also perturb the latching map via ac-Stark fluctuations and photon shot noise, motivating fast cavity reset and engineered readout modes \cite{sunada2024photon,swiadek2024enhancing,zhao2025single}. In our latching setting this backaction primarily affects fidelity through latch stability and resonator discrimination, because the classical latch outcome is fixed once $t_{\mathrm{lat}}$ has completed.

These considerations affect the two measurement protocols differently. In the sequential protocol, the later-latched degree of freedom undergoes additional idle evolution before its own $t_{\mathrm{lat}}$, increasing exposure to relaxation and dephasing. The simultaneous protocol, by contrast, features overlapping latching windows that minimize pre-latch idle time but heighten sensitivity to correlated noise and calibration drifts during the shared ramp.
Consequently, the predicted advantage of the energy basis in the single-qubit system and dressed basis in the coupled-qubit system is expected to persist when $t_{\mathrm{lat}} \ll T_1$ and $t_{\mathrm{lat}} \ll T_{\phi}^{(\mathrm{ramp})}$ and the latched QFP state remains stable throughout $t_{\mathrm{int}}$, where $T_{\phi}^{(\mathrm{ramp})}$ is the effective dephasing time during the annealing ramp. For the coupled-qubit case, an additional requirement is that the dressed-state splitting during latching exceeds the noise-induced broadening on timescale $t_{\mathrm{lat}}$, thereby suppressing transitions between dressed eigenstates. When noise is strong on the latching timescale or when the ramp transiently increases flux sensitivity, threshold broadening can wash out basis-dependent differences. In the strongly noise-dominated limit, the empirically optimal basis can become device- and operating-point dependent.
A quantitative determination of these crossovers requires an open-system model that combines a Lindblad description for photon loss and measurement backaction with a stochastic description of low-frequency flux noise and an effective model for QFP switching \cite{breuer2007oqs,clerk2010introduction,blais2021circuit,gambetta2008quantum}, which we leave for future work.

\section{Conclusions \label{sec:conclu}}

We perform a systematic theoretical and numerical investigation of QFP-mediated measurements of flux qubits, focusing on the interplay between state fidelity, measurement basis, and system parameters, such as measurement time, resonator coherent state amplitude, and qubit-qubit coupling strength. Using the QFP as an intermediary to latch and map the qubit state to a measurable quantity in the resonator, our approach provides a versatile framework for optimizing fidelity. This work offers two significant contributions: (i) a comprehensive comparison between energy and flux bases for readouts, demonstrating that the energy basis often yields higher fidelity in a wide range of scenarios, and (ii) an in-depth analysis of how QFP annealing processes affect fidelity and the selection of optimal readout strategies.

Our results highlight the QFP's ability to enable high-fidelity readout via adiabatic annealing, effectively latching the qubit state and enabling an effective flux qubit Hamiltonian description after annealing. We show that this process establishes entanglement between the qubit and QFP in both flux and energy bases, providing a robust means of transferring the state information into a form amenable to dispersive readout. Critically, our comparison of single-qubit readouts in different bases supports theoretical predictions that the energy basis generally outperforms the flux basis, underscoring the importance of basis selection for practical quantum readout protocols.
The oscillatory and nonmonotonic fidelity trends arise from photon-number conditioned phase accumulation during discrimination together with non-QND hybridization when transverse coupling components are present, leading to basis-dependent trade-offs and crossovers.

Extending beyond the single-qubit scenario, we examine the measurement of a target qubit in a two-qubit system and compare two distinct proposals: sequential and simultaneous measurements. The results show that sequential measurement consistently achieves high fidelity in energy bases and demonstrates greater resilience to qubit coupling than flux bases. However, this robustness is offset by slower measurement times, as fidelity degrades at very short measurement durations.
Conversely, the simultaneous readout model utilizes QFP annealing for both qubits, enabling high fidelity at shorter timescales and offering a pathway to rapid and efficient measurements. When transverse coupling is appreciable, noncommuting interactions drive coherent hybridization during discrimination, so that the effective measured observable crosses over from the bare basis toward the dressed basis as the integration time increases. Consequently, while simultaneous readout supports faster discrimination, it imposes stricter synchronization and timing constraints to capture the bare-basis projection before hybridization degrades the signal. These findings provide a guiding framework for the design of scalable, high-fidelity readout architectures in multi-qubit circuits.

Future research will focus on extending these analyses to open quantum systems and incorporating realistic noise and decoherence models. This will be essential to bridge the gap between idealized theoretical predictions and the performance of real superconducting qubit devices.
In parallel, it will be valuable to extend algorithm-hardware codesign \cite{ji2025algorithm,ji2024improving,ji2023optimizing,ji2024synergistic} to an algorithm-hardware-measurement codesign loop by optimizing the calibrated measurement channel via readout pulse shaping, integration time, and discrimination \cite{swiadek2024enhancing,bengtsson2024model,chatterjee2025enhanced} and then feeding the resulting measurement fidelity back into near-term compilation and qubit-mapping pipelines as a hardware-aware cost \cite{seif2024suppressing,khandavilli2023towards,zhu2025quantum}.
Finally, scaling these QFP-mediated readout techniques to larger qubit arrays and more complex multi-QFP architectures will establish foundational design principles for measurement optimization in next-generation superconducting processors.
By systematically integrating device design, basis engineering, and measurement strategy, this line of research advances the development of robust, high-fidelity qubit readout, an essential prerequisite for scalable and practical quantum computation.

\begin{acknowledgments}

The authors would like to thank Uwe Hartmann and Marius Schöndorf for their useful discussions.

\end{acknowledgments}

\appendix

\section{Dispersive readout and basis transformations}

\subsection{Rotating wave approximation and dressed states}\label{app:jc_dressed}

This section derives the Jaynes–Cummings Hamiltonian and establishes the bare and dressed basis notation.
We transform $\hat{H}_{\text{Rabi}}$ (Eq.~\eqref{equ:H_rabi}) into a rotating frame by applying a unitary transformation
\begin{equation*}
    \hat{U}_r=\exp(i\omega_{r} t \hat{a}^\dagger \hat{a}-i\omega_q t \hat{\sigma}_z /2),
\end{equation*}
yielding
\begin{equation*}
    \tilde{\hat{H}}_{\text{Rabi}}=\hat{U}_r\hat{H}_{\text{Rabi}}\hat{U}_r^\dagger+i\dot{\hat{U}}_r\hat{U}_r^\dagger.
\end{equation*}
Using the Baker-Campbell-Hausdorff formula, we obtain
\begin{align*}
    \tilde{\hat{H}}_{\text{Rabi}}= g\,\bigl[&\hat{a}\hat{\sigma}_{+}e^{i(\omega_q-\omega_r)}+\hat{a}^\dagger\hat{\sigma}_{-}e^{i(\omega_r-\omega_q)}\\
    +\,&\hat{a}\hat{\sigma}_{-}e^{-i(\omega_q+\omega_r)}+\hat{a}^\dagger\hat{\sigma}_{+}e^{i(\omega_q+\omega_r)}\bigr].
\end{align*}
Under the rotating wave approximation (RWA), where $\left(\omega_q + \omega_r\right) \gg \left\{g, \abs{\omega_q - \omega_r}\right\}$, two fast oscillating terms can be neglected. Transforming back into the Schrödinger picture, the system is described by the well-known Jaynes-Cummings Hamiltonian (Eq.~\eqref{equ:jaynes-cummings}).
Diagonalizing it in the subspace $\{\ket{g,n+1},~ \ket{e,n}\}$ yields the eigenvalues
\begin{equation*}
    E_{\pm,n}=\omega_r \left(n+\frac{1}{2}\right)\pm \frac{1}{2}\sqrt{\delta^2+4g^2\left(n+1\right)},
\end{equation*}
where $\delta=\omega_q-\omega_r$ is the detuning between the QFP and resonator frequencies. The corresponding eigenstates are
\begin{equation*}
\begin{split}
    \ket{+,n}&=\phantom{-}\cos(\frac{\theta_{n}}{2})\ket{e,n}+\sin(\frac{\theta_n}{2})\ket{g,n+1},\\
    \ket{-,n}&=-\sin(\frac{\theta_{n}}{2})\ket{e,n}+\cos(\frac{\theta_n}{2})\ket{g,n+1},
\end{split}
    \label{equ:dressedstate}
\end{equation*}
where the mixing angle $\theta_n$ is defined as $\tan(\theta_n)=2g \sqrt{n+1}/\delta$.
These eigenstates are referred to as \textit{dressed states} in quantum optics and atomic physics, while $\ket{g,n+1}$ and $\ket{e,n}$ are \textit{bare states}.

\subsection{Energy and flux basis transformations}\label{app:single_qubit_disp_steps}

Here we rotate the single-qubit readout Hamiltonian, apply displacement and dispersive transformations, and derive the effective Hamiltonian 
Eq.~\eqref{equ:H_3} used for single-qubit readout analysis. Transforming the Hamiltonian in Eq.~\eqref{eq:H_singleq_qfp} describing single-qubit readout into the qubit energy eigenbasis yields
\begin{eqnarray*}
\tilde{\hat{H}}&=&-\frac{ \omega_{\text{eff}}}{2}\tilde{\hat{\sigma}}_z + g\left[\cos(\theta_{\text{eff}}) \tilde{\hat{\sigma}}_z- \sin(\theta_{\text{eff}}) \tilde{\hat{\sigma}}_x\right] \left(\hat{a}^\dagger+\hat{a}\right)\nonumber\\
&& + \omega_r \hat{a}^\dagger\hat{a},
\end{eqnarray*}
where $\tilde{\hat{\sigma}}_z$ is the Pauli Z operator in the qubit eigenbasis, $\omega_{\text{eff}}=\sqrt{\epsilon_q^2+\Delta^2_{\text{eff}}}$, and $\theta_{\text{eff}}$ is defined as $\tan (\theta_{\text{eff}}) =\Delta_{\text{eff}}/\epsilon_q$.
Under the RWA, the Hamiltonian becomes
\begin{eqnarray*}
    \tilde{\hat{H}}&=&-\frac{\omega_{\text{eff}}}{2} \tilde{\hat{\sigma}}_z+ \omega_r \hat{a}^\dagger\hat{a}+ g  \cos(\theta_{\text{eff}})  \tilde{\hat{\sigma}}_z \left(\hat{a}^\dagger+\hat{a}\right)\nonumber\\
    & &- g  \sin(\theta_{\text{eff}})  \left(\hat{a}\tilde{\hat{\sigma}}_+ +\hat{a}^\dagger\tilde{\hat{\sigma}}_-\right).
    \label{equ:H_t}
\end{eqnarray*}
Applying the displacement operator 
$\hat{D}(\alpha)=\exp(\alpha \hat{a}^{\dag} -\alpha^* \hat{a})$ and choosing $\alpha=-g\cos(\theta_{\text{eff}})\tilde{\hat{\sigma}}_z /\omega_r$, we obtain
\begin{equation*}
    \tilde{\hat{H}}=-\frac{\omega_{\text{eff}}}{2} \tilde{\hat{\sigma}}_z+\omega_r \hat{a}^{\dag}\hat{a}- g  \sin(\theta_{\text{eff}})\left(a \tilde{\hat{\sigma}}_+ + \hat{a}^{\dag} \tilde{\hat{\sigma}}_-\right).
    \label{equ:Htilde1}
\end{equation*}
In the dispersive regime with $\delta = \omega_{\text{eff}} - \omega_r  \neq 0$ and $g  \sin(\theta_{\text{eff}}) \ll \abs{\delta}$, the Hamiltonian becomes
\begin{equation*}
    \tilde{\hat{H}}=-\frac{\delta}{2} \tilde{\hat{\sigma}}_z- g  \sin(\theta_{\text{eff}})\left(a \tilde{\hat{\sigma}}_+ + \hat{a}^{\dag} \tilde{\hat{\sigma}}_-\right).
    \label{equ:H2}
\end{equation*}
Finally, applying a unitary transformation $\hat{U}_3=\exp(\lambda\left(a\tilde{\hat{\sigma}}_+ - \hat{a}^{\dag}\tilde{\hat{\sigma}}_- \right))$ with $\lambda=g\sin(\theta_{\text{eff}})/\delta$, yields the Hamiltonian in the energy basis (Eq.~\eqref{equ:H_3}) up to the first order in $\lambda$.

\section{Backaction of QFP annealing on flux qubits \label{sec:appen}}

This section investigates the backaction effects of QFP annealing on flux qubits. We focus on deriving the effective Hamiltonian (Eq.~\eqref{eq:effec_flux_qubit_hamil}) that governs the coupled system of the flux qubit and QFP after the annealing process. Additionally, we analyze how QFP annealing affects the state of the flux qubit connected to the QFP and the coupling strength $J$ (Eqs.~\eqref{eq:hamil_seque_measu} and \eqref{eq:hamil_simul_measu}) between coupled flux qubits. 

\subsection{Effective Hamiltonian of the coupled flux qubit and QFP system \label{subsec:effec_hamil_coupl_flux_qfp}}

We begin from the Hamiltonian that describes the QFP and its interaction with the flux qubit in Eq.~\eqref{equ:hamiltonian_qubit_qfp}
\begin{equation}
    \hat{H}_{\text{qfp}}(t) =  \tilde{\Omega}(t) \left[\hat{a}^\dagger \hat{a} + \nu(t) \left(\hat{a}^\dagger+\hat{a}\right) \hat{\sigma}_z\right],
    \label{equ:hamiltonian_qfp_int}
\end{equation}
where
\begin{equation*}
\nu(t)\equiv \frac{\tilde g(t)\varphi_p(t)}{\tilde\Omega(t)}
=\varphi_p(t)\sqrt{\frac{m\,\tilde\Omega(t)}{2E_L}}
\end{equation*}
is the displacement amplitude in oscillator units assumed real and $\varphi_p(t)$ denotes the time-dependent position of the instantaneous QFP potential minimum during the annealing schedule. Eq.~\eqref{equ:hamiltonian_qfp_int} is a linearly displaced harmonic oscillator
\begin{eqnarray*}
\hat{H}_{\text{qfp}}(t)&=&\tilde{\Omega}(t)\left[\hat{a}^\dagger\hat{a}+\hat{\sigma}_z\,\nu(t)\left(\hat{a}^\dagger+\hat{a}\right)\right]\\
&=&\tilde{\Omega}(t)\left[\left(\hat{a}^\dagger+\hat{\sigma}_z\nu\right)\left(\hat{a}+\hat{\sigma}_z\nu\right)-\nu^2\right].
\end{eqnarray*}
Defining $\hat{U}(t)=\hat{D}\!\left[-\hat{\sigma}_z\nu(t)\right]$ with the displacement operator $\hat{D}(\alpha)=\exp(\alpha \hat{a}^\dagger - \alpha^*\hat{a})$, we have
\begin{equation*}
    \left(\hat{a}^\dagger + \hat{\sigma}_z \nu\right) \left( \hat{a} + \hat{\sigma}_z \nu\right) = \hat{U}(t)\hat{a}^\dagger\hat{a}\hat{U}^\dagger(t),
\end{equation*}
yielding the instantaneous eigenstates and eigenvalues
\begin{eqnarray*}
\ket{M_\pm(t)}&=&\hat D[-\hat{\sigma}_z\nu(t)]\ket{M},\\
E_M(t)&=&\tilde{\Omega}(t)\left(M-\nu^2(t)\right),
\end{eqnarray*}
where $\ket{M}$ are Fock states with $M=0,1,2,\dots$ and we have dropped the usual zero-point constant consistently.
In the displaced-oscillator frame defined by
$\ket{\tilde\psi(t)}=\hat U^\dagger(t)\ket{\psi(t)}$, the Hamiltonian acquires the standard inertial term
\begin{equation}
    \tilde{\hat{H}}_{\text{qfp}}(t)
    = \hat{U}^\dagger(t)\hat{H}_{\text{qfp}}(t)\hat{U}(t)
    - i\,\hat{U}^\dagger(t)\dot{\hat{U}}(t),
    \label{eq:Hqfp_displaced_frame}
\end{equation}
where
\begin{equation}
-i\,\hat{U}^\dagger(t)\dot{\hat{U}}(t)
= i\,\dot{\nu}(t)\,\hat{\sigma}_z\left(\hat a^\dagger-\hat a\right).
\label{eq:inertia_term}
\end{equation}
Additional nonadiabatic terms arise if $\tilde\Omega(t)$ varies appreciably, which are suppressed under Eq.~\eqref{eq:adiabatic_pointer}.
This transformation also affects the flux qubit Hamiltonian $\hat{H}_{q}$, given by
\begin{equation}
    \hat{H}_{q} = -\frac{1}{2}\left(\epsilon_q \hat{\sigma}_z + \Delta_q \hat{\sigma}_x\right),
\end{equation}
leading to
\begin{eqnarray}
    \tilde{\hat{H}}_{q} &=& \hat{U}^\dagger(t) \hat{H}_{q} \hat{U}(t)\nonumber\\
    &=& \hat{D}\left(\hat{\sigma}_z \nu\right) \hat{H}_{q} \hat{D}^\dagger\left(\hat{\sigma}_z \nu\right).
    \label{eq:bacha_flux_qubit}
\end{eqnarray}
To identify the leading inelastic correction in oscillator quadratures, we expand to first order and obtain
\begin{eqnarray}
    \hat{D}\left(\hat{\sigma}_z \nu\right) \hat{\sigma}_z \hat{D}^\dagger\left(\hat{\sigma}_z \nu\right) &=& \hat{\sigma}_z,\\
    \hat{D}\left(\hat{\sigma}_z \nu\right) \hat{\sigma}_x \hat{D}^\dagger\left(\hat{\sigma}_z \nu\right) &\approx& \hat{\sigma}_x + i\hat{\sigma}_y[2\nu(\hat{a}^\dagger - \hat{a})]\nonumber\\
    &=& \hat{\sigma}_x + 2\varphi_p\hat{q}\hat{\sigma}_y,
\end{eqnarray}
where
\begin{equation*}
\hat{q}=i\sqrt{\frac{m\tilde\Omega}{2E_L}}\left(\hat{a}^\dagger-\hat{a}\right)
\qquad \text{and} \qquad
\delta\hat\varphi=\sqrt{\frac{E_L}{2m\tilde\Omega}}\left(\hat a^\dagger+\hat a\right)
\end{equation*}
are canonically conjugate QFP quadratures, as defined in Sec.~\ref{subsec:qfp}.
In this expansion, the correction $2\varphi_p\hat q\,\hat\sigma_y$ is the leading inelastic term. It is off-diagonal in the QFP excitation number and therefore contributes to the low-energy dynamics only via transitions out of the instantaneous displaced manifold. In the adiabatic-pointer regime, where the QFP follows its instantaneous displaced ground state during the ramp, such manifold-changing transitions are suppressed dynamically, by the inertia term \eqref{eq:inertia_term} and energetically, by the large QFP level spacing $\tilde\Omega(t)$.
A sufficient set of adiabatic conditions is \cite{kato1950adiabatic,teufel2003adiabatic,bravyi2011schrieffer}
\begin{equation}
\left|\dot\nu(t)\right|\ll \tilde\Omega(t)
\qquad \text{and} \qquad
\left|\dot{\tilde\Omega}(t)\right|\ll \tilde\Omega^2(t),
\label{eq:adiabatic_pointer}
\end{equation}
which ensure that the inertial and curvature-driven couplings do not populate excited QFP levels. In addition, the inelastic tunneling channel generated by the  $\hat q$ term is perturbative when
\begin{equation}
\Delta_q\,|\varphi_p(t)|\,\Big|\langle 1|\hat q|0\rangle\Big|
=\Delta_q\,|\nu(t)|
\ll \tilde\Omega(t),
\label{eq:adiabatic_inelastic}
\end{equation}
so that the $  \hat q  $ term only produces virtual admixtures of excited displaced-oscillator states with relative weight of order $(\Delta_q\nu/\tilde\Omega)^2$ and can be neglected to leading order. Under this approximation, we neglect manifold-changing matrix elements that couple different displaced-oscillator levels and Eq.~\eqref{equ:hamiltonian_qubit_qfp} simplifies to \cite{twyeffort2005dynamics, schondorf2020flux}
\begin{equation}
    \hat{H} = \bigoplus_{M=0}^{\infty} \hat{H}_M,
\end{equation}
where $\hat{H}_M$ is the Hamiltonian in the $M$-excitation subspace $\{\ket{0, M_{+}}, \ket{1, M_{-}}\}$ and is expressed as
\begin{equation}
    \hat{H}_M = \begin{bmatrix}
        E_M-\epsilon_q/2 & -\Delta_q\bra{M_{-}}\ket{M_{+}}/2\\
        -\Delta_q\bra{M_{-}}\ket{M_{+}}/2 & E_M+\epsilon_q/2\\
    \end{bmatrix},\\
\end{equation}
or equivalently
\begin{equation}
    \hat{H}_M = E_M \mathbb{1} - \frac{\epsilon_q}{2} \hat{\sigma}_z - \frac{\Delta_{\text{eff}}}{2} \hat{\sigma}_x
    \label{eq:effec_hamilt_M}
\end{equation}
with $\Delta_{\text{eff}}=\Delta_q \bra{M_{-}}\ket{M_{+}}$. The overlap $\bra{M_{-}}\ket{M_{+}}$ is given by \cite{twyeffort2005dynamics}
\begin{equation*}
    \bra{M_{-}}\ket{M_{+}} = e^{-\eta} L_M\left(2 \eta\right),
\end{equation*}
where $\eta=2\nu^2$ and $L_M$ is the Laguerre polynomial.
Assuming the harmonic oscillator is in its ground state ($M=0$), the effective Hamiltonian in Eq.~\eqref{eq:effec_flux_qubit_hamil} is obtained by neglecting the constant term in Eq.~\eqref{eq:effec_hamilt_M}. In this case, $\Delta_{\text{eff}}$ is given by
\begin{equation}
    \Delta_{\text{eff}}=\Delta_q\bra{0_{-}}\ket{0_{+}} = \Delta_q e^{-\eta}.
\end{equation}

\subsection{Backaction of QFP annealing on coupled flux qubits \label{subsec:backa_qfp_flux_qub}}

Here we demonstrate that the coupling strength $J$ between two flux qubits in Eq.~\eqref{eq:hamil_seque_measu} remains unaffected during the QFP annealing process.
The interaction Hamiltonian describing two coupled flux qubits is given by
\begin{equation}
    \hat{H}_{I} = J\hat{\sigma}_1^z\hat{\sigma}_2^z.
\end{equation}
Analogous to the analysis in Eq.~\eqref{eq:bacha_flux_qubit}, the backaction of QFP annealing can be evaluated by applying the displacement operator to $\hat{H}_{I}$
\begin{equation}
    \tilde{\hat{H}}_{I} = \hat{D}\left(\hat{\sigma}_2^z \nu\right) \hat{H}_{I} \hat{D}^\dagger\left(\hat{\sigma}_2^z \nu\right).
\end{equation}
Applying this transformation yields
\begin{eqnarray}
    \tilde{\hat{H}}_{I} &=& J\hat{\sigma}_1^z \hat{D}\left(\hat{\sigma}_2^z \nu \right) \hat{\sigma}_2^z \hat{D}^\dagger\left(\hat{\sigma}_2^z \nu\right),\\
    &=& J\hat{\sigma}_1^z\hat{\sigma}_2^z = \hat{H}_{I}.
\end{eqnarray}
This result indicates that the interaction Hamiltonian $\hat{H}_{I}$ remains unchanged after the transformation, implying that the coupling strength $J$ between the two flux qubits is unaffected by the QFP annealing process.
Moreover, since the displacement during annealing is conditioned on the target qubit $2$, $\hat U_2(t)=\hat D[-\hat\sigma_2^z\nu(t)]$, and $[\hat H_I,\hat\sigma_2^z]=0$, it follows that $[\hat{H}_I,\hat{U}_2(t)]=0$. Therefore the inertial term $-i\,\hat U_2^\dagger(t)\dot{\hat U}_2(t)$ does not modify $\hat{H}_I$ and therefore does not renormalize $J$.
Similarly, the coupling strength $J$ in Eq.~\eqref{eq:hamil_simul_measu} is also not renormalized by the QFP annealing. The consistency of $J$ before and after the annealing process ensures that the qubit-qubit interactions are preserved, which is crucial for maintaining the integrity of quantum operations in coupled qubit systems.

\section{Multi-qubit basis transformations}

Here we collect the basis rotations and diagonalizations used in Sec.~\ref{sec:measu} to relate bare and dressed measurement bases for the sequential and simultaneous two-qubit protocols.

\subsection{Sequential model in $q_2$ energy basis}\label{app:seq_q2_dress}

In the rotating frame with frequency $\omega_r$, the Hamiltonian in Eq.~\eqref{eq:hamil_seque_measu} describing sequential measurement in the flux basis is expressed in the $q_2$ energy basis as
\begin{eqnarray}
    \tilde{\hat{H}}&=&-\frac{\delta_{\text{eff,2}}}{2}\tilde{\hat{\sigma}}_2^z+J\hat{\sigma}_1^z\left[\cos(\theta_{\text{eff,2}}) \tilde{\hat{\sigma}}_2^z-\sin(\theta_{\text{eff,2}}) \tilde{\hat{\sigma}}_2^x\right] \nonumber\\
    & &+g\left[\cos(\theta_{\text{eff,2}}) \tilde{\hat{\sigma}}_2^z-\sin(\theta_{\text{eff,2}}) \tilde{\hat{\sigma}}_2^x\right]\left(\hat{a}^{\dag}+\hat{a}\right),
\end{eqnarray}
where $\theta_{\text{eff,2}}$ is defined as $\tan(\theta_{\text{eff,2}})=\Delta_{\text{eff,2}}/\epsilon_{2}$ and $\delta_{\text{eff,2}}=\omega_{\text{eff,2}}-\omega_r$ is effective frequency detuning with $\omega_{\text{eff,2}}=\sqrt{\epsilon_{2}^2+\Delta_{\text{eff,2}}^2}$. Applying the displacement operator $\hat{D}(\alpha)$ with $\alpha=-g\cos(\theta_{\text{eff,2}})\tilde{\hat{\sigma}}_2^z/\omega_r$, we obtain the Hamiltonian under the RWA as
\begin{eqnarray}
        \tilde{\hat{H}}&=&-\frac{\delta_{\text{eff,2}}}{2}\tilde{\hat{\sigma}}_2^z+J\hat{\sigma}_1^z\left[\cos(\theta_{\text{eff,2}}) \tilde{\hat{\sigma}}_2^z-\sin(\theta_{\text{eff,2}}) \tilde{\hat{\sigma}}_2^x\right]\nonumber\\
        & &-g\sin(\theta_{\text{eff,2}}) \left(\hat{a}^{\dag}\tilde{\hat{\sigma}}_2^-+\hat{a}\tilde{\hat{\sigma}}_2^+\right).
    \label{equ:Hq2}
\end{eqnarray}
To eliminate the interaction term between $q_2$ and resonator, we apply a unitary operator $\hat{U}=\exp\left(\lambda\left(\hat{a}\tilde{\hat{\sigma}}_2^+-\hat{a}^{\dag}\tilde{\hat{\sigma}}_2^-\right)\right)$ with $\lambda=g\sin(\theta_{\text{eff,2}})/\delta_{\text{eff,2}}$, to Eq.~\eqref{equ:Hq2}, yielding Eq.~\eqref{equ:Ht2_q2eig} in the main text.

Diagonalization of Hamiltonian described in Eq.~\eqref{equ:Ht2_q2eig} yields the eigenvalues
\begin{equation*}
    E_{(e_2)}=\big\{ -\omega_{n-},~
    \omega_{n-},~-\omega_{n+},
    ~\omega_{n+} \big\},
\end{equation*}
where $\omega_{n\pm}=\sqrt{\left(\delta_{\text{eff2, n}}/2\pm J_{zz}\right)^2+J_{zx}^2}$ with $J_{zz} = J \cos(\theta_{\text{eff,2}})$ and $J_{zx} = J \sin(\theta_{\text{eff,2}})$. The eigenstates are given by
\begin{eqnarray*}
    \ket{\overline{00}}&=&\phantom{-}\cos(\frac{\theta_{n-}}{2})\ket{\tilde 0\tilde 0}+\sin(\frac{\theta_{n-}}{2})\ket{\tilde 0\tilde 1},\nonumber\\
    \ket{\overline{01}}&=&-\sin(\frac{\theta_{n-}}{2})\ket{\tilde 0\tilde 0}+\cos(\frac{\theta_{n-}}{2})\ket{\tilde 0\tilde 1},\nonumber\\
    \ket{\overline{10}}&=&\phantom{-}\cos(\frac{\theta_{n+}}{2})\ket{\tilde1\tilde 0}+\sin(\frac{\theta_{n+}}{2})\ket{\tilde1\tilde 1}, \nonumber\\
    \ket{\overline{11}}&=&-\sin(\frac{\theta_{n+}}{2})\ket{\tilde1\tilde 0}+\cos(\frac{\theta_{n+}}{2})\ket{\tilde1\tilde 1},
\end{eqnarray*}
with $\theta_{n\pm}$ satisfying $\tan(\theta_{n\pm})=\mp J_{zx}/\left(\delta_{\text{eff2,n}}/2 \pm J_{zz}\right)$.
The relationship between the bare and dressed bases is described by
\begin{equation*}
        \begin{bmatrix}
            \hat{\sigma}_2^z+\hat{\sigma}_1^z\tilde{\hat{\sigma}}_2^z\\
            \hat{\sigma}_{2}^{x}+\hat{\sigma}_{1}^{z}\tilde{\hat{\sigma}}_2^x\\
            \hat{\sigma}_2^z-\hat{\sigma}_1^z\tilde{\hat{\sigma}}_2^z\\
            \hat{\sigma}_{2}^{x}-\hat{\sigma}_{1}^{z}\tilde{\hat{\sigma}}_2^x
        \end{bmatrix}=A\begin{bmatrix}
            \bar{\hat{\sigma}}_{2}^{z}+\bar{\hat{\sigma}}_{1}^{z}\bar{\hat{\sigma}}_{2}^{z}\\
            \bar{\hat{\sigma}}_{2}^{x}+\bar{\hat{\sigma}}_{1}^{z}\bar{\hat{\sigma}}_{2}^{x}\\
            \bar{\hat{\sigma}}_{2}^{z}-\bar{\hat{\sigma}}_{1}^{z}\bar{\hat{\sigma}}_{2}^{z}\\
            \bar{\hat{\sigma}}_{2}^{x}-\bar{\hat{\sigma}}_{1}^{z}\bar{\hat{\sigma}}_{2}^{x}
        \end{bmatrix}
\end{equation*}
with
\begin{equation*}
    A = \begin{bmatrix}
        \cos(\theta_{0-}) & -\sin(\theta_{0-}) & 0 & \phantom{-}0\\
        \sin(\theta_{0-}) & \phantom{-}\cos(\theta_{0-}) & 0 & \phantom{-}0\\
        0 & \phantom{-}0 & \cos(\theta_{0+}) & -\sin(\theta_{0+})\\
        0 & \phantom{-}0 & \sin(\theta_{0+}) & \phantom{-}\cos(\theta_{0+})
    \end{bmatrix} .
\end{equation*}

Using these transformation relations, we obtain the Hamiltonian in the dressed basis
\begin{eqnarray*}
    \bar{\hat{H}}^{(2)}&=&-\frac{\omega_{n-}}{2}\left[\cos(\theta_{n-}-\theta_{0-})\left(\bar{\hat{\sigma}}_{2}^{z}+\bar{\hat{\sigma}}_{1}^{z}\bar{\hat{\sigma}}_{2}^{z}\right)\right.\nonumber\\
    &&\qquad\left.+ \sin(\theta_{n-}-\theta_{0-}) \left(\bar{\hat{\sigma}}_{2}^{x}+\bar{\hat{\sigma}}_{1}^{z}\bar{\hat{\sigma}}_{2}^{x}\right)\right] \nonumber\\
    &&-\frac{\omega_{n+}}{2} \left[\cos(\theta_{n+}-\theta_{0+}) \left(\bar{\hat{\sigma}}_{2}^{z}-\bar{\hat{\sigma}}_{1}^{z}\bar{\hat{\sigma}}_{2}^{z}\right)\right.\nonumber\\
    &&\qquad\left.+\sin(\theta_{n+}-\theta_{0+}) \left(\bar{\hat{\sigma}}_{2}^{x}-\bar{\hat{\sigma}}_{1}^{z}\bar{\hat{\sigma}}_{2}^{x}\right)\right].
    \label{equ:h2d}
\end{eqnarray*}

\subsection{Sequential model in $q1$-$q2$ energy basis}\label{app:seq_q1_q2_xx}

Diagonalizing the xx-approximation Hamiltonian $\tilde{\hat{H}}^{(e)}_{xx}$ in Eq.~(\ref{equ:Hq4_xx}) yields the eigenstates
\begin{eqnarray*}
    \ket{\overline{00}}&=&\phantom{-}\cos(\frac{\theta_{0+}}{2})\ket{\tilde0\tilde 0}+\sin(\frac{\theta_{0+}}{2})\ket{\tilde1\tilde 1}, \nonumber\\
    \ket{\overline{01}}&=&\phantom{-}\cos(\frac{\theta_{0-}}{2})\ket{\tilde0\tilde 1}+\sin(\frac{\theta_{0-}}{2})\ket{\tilde1\tilde 0}, \nonumber\\
    \ket{\overline{10}}&=&-\sin(\frac{\theta_{0-}}{2})\ket{\tilde0\tilde 1}+\cos(\frac{\theta_{0-}}{2})\ket{\tilde1\tilde 0}, \nonumber\\
    \ket{\overline{11}}&=&-\sin(\frac{\theta_{0+}}{2})\ket{\tilde0\tilde 0}+\cos(\frac{\theta_{0+}}{2})\ket{\tilde1\tilde 1},
\end{eqnarray*}
where $\theta_{n\pm }$ is defined as $\tan(\theta_{n\pm })=\pm2J_{xx}/\delta_{\text{eff2,n}}$. The corresponding eigenvalues are
\begin{equation*}
E^{(e)}_{xx}= \{-\omega_n, ~ -\omega_n, ~\omega_n, ~ \omega_n\},
\end{equation*}
where $\omega_n = \sqrt{J_{xx}^2+\left(\delta_{\text{eff2,n}}/2\right)^2}$.
This allows us to represent the Hamiltonian in the dressed basis
\begin{align}
    \bar{H}_{xx}^{(e)}=\frac{\omega_n}{2}\,\bigl[
    &\cos(\theta_{n-}-\theta_{0-}) \left(\bar{\hat{\sigma}}_{1}^{z}+\bar{\hat{\sigma}}_{2}^{z}\right)\nonumber\\
    +\,&\sin(\theta_{n-}-\theta_{0-}) \left(\bar{\hat{\sigma}}_{1}^{x}\bar{\hat{\sigma}}_{2}^{x}-\bar{\hat{\sigma}}_{1}^{y}\bar{\hat{\sigma}}_{2}^{y}\right)\nonumber\\
    +\,&\cos(\theta_{n+}-\theta_{0+})\left(\bar{\hat{\sigma}}_{1}^{z}-\bar{\hat{\sigma}}_{2}^{z}\right)\nonumber\\
    +\,&\sin(\theta_{n+}-\theta_{0+}) \left(\bar{\hat{\sigma}}_{1}^{x}\bar{\hat{\sigma}}_{2}^{x}+\bar{\hat{\sigma}}_{1}^{y}\bar{\hat{\sigma}}_{2}^{y}\right)\bigr].
    \label{equ:Ht2_xx_q1q2}
\end{align}

\subsection{Simultaneous measurement}\label{app:sim_q1_q2_energy}

The Hamiltonian describing simultaneous measurement in the flux basis, given by Eq.~\eqref{eq:hamil_simul_measu}, is expressed in the $q_1$-$q_2$ energy basis as follows
\begin{eqnarray}
    \tilde{\hat{H}}&=&-\frac{\omega_{\text{eff,1}}}{2}\tilde{\hat{\sigma}}_1^z+g_2\left[\cos(\theta_{\text{eff,2}}) \tilde{\hat{\sigma}}_2^z-\sin(\theta_{\text{eff,2}}) \tilde{\hat{\sigma}}_2^x\right]\left(\hat{a}^{\dag}+\hat{a}\right)\nonumber\\
    & &-\frac{\omega_{\text{eff,2}}}{2}\tilde{\hat{\sigma}}_2^z+\omega_r \hat{a}^{\dag}\hat{a}+J\left[\cos(\theta_{\text{eff,1}}) \tilde{\hat{\sigma}}_1^z-\sin(\theta_{\text{eff,1}}) \tilde{\hat{\sigma}}_1^x\right]\nonumber\\
    & & \otimes \left[\cos(\theta_{\text{eff,2}}) \tilde{\hat{\sigma}}_2^z-\sin(\theta_{\text{eff,2}}) \tilde{\hat{\sigma}}_2^x\right],
\label{equ:hfq1fq2}
\end{eqnarray}
where $\omega_{\text{eff,i}}=\sqrt{\epsilon_{i}^2+\Delta_{\text{eff,i}}^2}$ and $\theta_{\text{eff,i}}$ is defined as $\tan(\theta_{\text{eff,i}})=\Delta_{\text{eff,i}}/\epsilon_{i}$ with $i\in \{1,2\}$.
Applying the displacement operator $\hat{D}(\alpha)$ with $\alpha=-g_2\cos(\theta_{\text{eff,2}}) \tilde{\hat{\sigma}}_2^z/\omega_r$, we obtain the Hamiltonian under the RWA
\begin{eqnarray}
    \tilde{\hat{H}}&=&-\frac{\omega_{\text{eff,1}}}{2}\tilde{\hat{\sigma}}_1^z
    -g_2\sin(\theta_{\text{eff,2}}) \left(\hat{a}^{\dag}\tilde{\hat{\sigma}}_2^-+\hat{a}\tilde{\hat{\sigma}}_2^+\right)+\omega_r \hat{a}^{\dag}\hat{a}\nonumber\\
    &&-\frac{\omega_{\text{eff,2}}}{2}\tilde{\hat{\sigma}}_2^z+J\left[\cos(\theta_{\text{eff,1}}) \tilde{\hat{\sigma}}_1^z-\sin(\theta_{\text{eff,1}}) \tilde{\hat{\sigma}}_1^x\right]\nonumber\\
    && \otimes\left[\cos(\theta_{\text{eff,2}}) \tilde{\hat{\sigma}}_2^z-\sin(\theta_{\text{eff,2}}) \tilde{\hat{\sigma}}_2^x\right].
    \label{equ:H_3_q1q2}
\end{eqnarray}
Further eliminating the interaction term between qubit and resonator by applying the unitary operator $\hat{U}=\exp\left[\lambda_2\left(\hat{a}\tilde{\hat{\sigma}}_2^+-\hat{a}^{\dag}\tilde{\hat{\sigma}}_2^-\right)\right]$ with $\lambda_2=g_2\sin(\theta_{\text{eff,2}})/\Delta_{\text{eff,2}}$, we obtain
\begin{eqnarray}
    \tilde{\hat{H}}&=&-\frac{\omega_{\text{eff,1}}}{2}\tilde{\hat{\sigma}}_1^z
    -\left(\frac{\omega_{\text{eff,2}}+\chi}{2}+\chi\hat{a}^\dagger\hat{a}\right)\tilde{\hat{\sigma}}_2^z+\omega_r\hat{a}^\dagger\hat{a}\nonumber\\
    & &+J\left[\cos(\theta_{\text{eff,1}}) \tilde{\hat{\sigma}}_1^z-\sin(\theta_{\text{eff,1}}) \tilde{\hat{\sigma}}_1^x\right]\nonumber\\
    & & \otimes \left[\cos(\theta_{\text{eff,2}}) \tilde{\hat{\sigma}}_2^z-\sin(\theta_{\text{eff,2}}) \tilde{\hat{\sigma}}_2^x\right].
\end{eqnarray}
Finally, transforming to a frame rotating at the frequency of $\omega_r$ yields Eq.~\eqref{equ:ht3_q1q1_b} in the main text.

\bibliography{refs}

\end{document}